\RequirePackage{amsthm}
\documentclass[sn-mathphys-num]{sn-jnl}

\usepackage{graphicx}%
\usepackage{multirow}%
\usepackage{amsmath,amssymb,amsfonts}%
\usepackage{amsthm}%
\usepackage{mathrsfs}%
\usepackage[title]{appendix}%
\usepackage{xcolor}%
\usepackage{textcomp}%
\usepackage{manyfoot}%
\usepackage{booktabs}%
\usepackage{algorithm}%
\usepackage{algorithmicx}%
\usepackage{algpseudocode}%
\usepackage{listings}%

\theoremstyle{thmstyleone}%
\newtheorem{theorem}{Theorem}
\newtheorem{proposition}[theorem]{Proposition}%
\theoremstyle{thmstyletwo}%
\newtheorem{remark}{Remark}%
\theoremstyle{thmstylethree}%
\usepackage{anyfontsize}
\usepackage{cases}

\usepackage[T1]{fontenc}
\usepackage{HT_command}
\usepackage{mathtools}
\usepackage{enumerate}
\usepackage{tikz}
\usepackage{amsmath}
\usepackage{tikz}
\usepackage{mathdots}
\usepackage{cancel}
\usepackage{color}
\usepackage{siunitx}
\usepackage{array}
\usepackage{multirow}
\usepackage{amssymb}
\usepackage{tabularx}
\usepackage{pgfplots}
\usepgfplotslibrary{groupplots}
\usetikzlibrary{plotmarks}
\pgfplotsset{compat=1.18} 
\usepgfplotslibrary{patchplots}

\begin{document}

\title[flatnessRelu]{An ANN-Enhanced Approach for Flatness-Based Constrained Control of Nonlinear Systems}


\author*[1]{\fnm{Huu-Thinh} \sur{Do}}
\email{huu-thinh.do@lcis.grenoble-inp.fr}

\author[1]{\fnm{Ionela} \sur{Prodan}}\email{ionela.prodan@lcis.grenoble-inp.fr}

\author[2]{\fnm{Florin} \sur{Stoican}}\email{florin.stoican@upb.ro}

\affil[1]{\orgname{Univ. Grenoble Alpes, Grenoble INP$^\dagger$, LCIS} \orgaddress{, \city{Valence}, \postcode{F-26000}, \country{France.\\$\dagger$Institute of Engineering and Management University Grenoble Alpes}}}


\affil[2]{\orgdiv{Department of Automatic Control  and Systems Engineering}, \orgname{UPB}, \country{Romania}}


\abstract{
Neural networks have proven practical for a synergistic combination of advanced control techniques. This work analyzes the implementation of rectified linear unit neural networks to achieve constrained control in differentially flat systems. Specifically, the class of flat systems enjoys the benefit of feedback linearizability, i.e., the systems can be linearized by means of a proper variable transformation. However, linearizing the dynamics comes at the price of distorting the constraint descriptions. We show that, by using neural networks, these constraints can be approximated as a union of polytopes, enabling the use of mixed-integer programming tools to guarantee constraint satisfaction. We further analyze the integration of the characterization into efficient settings such as control Lyapunov function-based and model predictive control (MPC). Interestingly, this description also allows us to explicitly compute the solution of the MPC problem for the nonlinear system. Several examples are provided to illustrate the effectiveness of our framework. 
}




\maketitle

\section{Introduction}\label{sec:Intro}

The feedback linearization (FL) technique, although introduced for decades \cite{isidori1996feedback}, still receives significant attention from control practitioners. 
In the set of feedback linearizable systems, differentially flat systems constitute a pertinent subclass with diverse applications in robotics \cite{Haithem2012flapping,li2021quadruped}, aerospace \cite{farahani2016constrained,sanchez2020flatness}, chemical \cite{utz2012flatness,mounier1998flatness} or even biomedical engineering for COVID-19 treatment \cite{Christoph2022}.
The idea of flatness-based feedback linearization (FB-FL), or FL in general, is to find a coordinate system in which the nonlinear dynamics can be linearized in closed-loop.
The immediate benefit of this setting derives from the direct access to linear control theory tools, from simple loop-shaping \cite{do2022tracking,levine2009analysis}, to advanced techniques like sliding mode or Model Predictive Control (MPC) \cite{craneZhang2017,greeff2018flatness,kong2023stable}. Furthermore, and perhaps even more interestingly, having an equivalent linear representation also allows us to leverage the measurement data for identification, robustification and prediction  \cite{alsalti2023data,de2023data,hou2016overview}, using behavioral theory \cite{willems2005note}. 
However, these advantages, while significant, typically fall short under the presence of constraints.
The reason is that, although the nonlinearity is removed from the dynamics in the new linearizing space, the constraints themselves become distorted and vary depending on the system's current state \cite{greco2022approximate,kong2023stable,skogestad2023transformed}.


For several applications, this problem has been successfully addressed by using the constraint's particular properties and control techniques like nested or minimally invasive control, notably for inverted pendulums \cite{ibanez2008controlling}, quadcopters \cite{nguyen2020flat}, and wheeled mobile robots \cite{tiriolo2023set}. These solutions, although effective, are case-dependent.
To develop a more general formula, 
MPC, due to its constraint-handling capabilities, appears highly suitable to handle the distorted constraints. However, since the constraint in the linearizing space is now state dependent and hence time-varying, it is required to optimize also such constraints over the prediction horizon, significantly increasing the online computational burden. To sidestep this problem, one of the simplest ``tricks'' is to derive the constraint set point-wise in time with feedback, and use it constantly over the prediction horizon \cite{skogestad2023transformed,nevistic1994feasible}. 
In this way, the online optimization problem becomes tractable. However, since the constraint is only considered for the current state, the forecasted trajectory may not be realizable, posing challenges in proving the feasibility and stability for such a scheme \cite{kurtz1998feedback}.

Another {approach} is to obtain an inner approximation of the distorted constraints which {can be efficiently managed by solvers, ensuring continuous constraint satisfaction while keeping the system's trajectory inside a tightened feasible domain.} For instance, in \cite{hall2023differentially,greeff2021learning}, Gaussian Processes were used to learn the linearizing mappings, and hence the propagated constraints, making the online MPC a second-order cone program. 

In our previous work
\cite{do2024_reluANN}, we demonstrated that rectified linear unit (ReLU) artificial neural networks (ANN) can also be used to {effectively} approximate the constraints. Furthermore, thanks to the piecewise affine activation function, the approximated constraints can be represented {by 
mixed-integer (MI) linear encodings. With this representation, the constraints can be taken into account for both the applied input and the forecasted trajectory.}
As a result, stabilizing admissible controllers can be obtained by employing such characterizations in optimization-based frameworks such as Control Lyapunov function-based control or MPC. 


In contrast, this work extends beyond Mixed-integer Program (MIP) representations and explores a geometric perspective on constrained FB-FL control. 
\thinh{Specifically, in the context of constrained FL-based control, we introduce:}
\begin{thinhnote}
\begin{itemize}
    \item an algorithm to derive polytopic partitions representing the constraints in the linearizing space;
    \item integration of the deduced piecewise-affine (PWA) constraints into optimization-based control frameworks, including CLF-based control and MPC;
    \item a proof that constraints in the linearizing space can be represented as unions of polytopes, rather than relying solely on MI formulations. This allows the use of efficient set-theoretic tools to develop constraint-compatible policies;
    \item an application of multi-parametric programming \cite{kvasnica2004multi} to compute the explicit solution of the MPC problem for the nonlinear system, facilitating the detection of feasible domains for optimization-based controllers.
\end{itemize}
    
\end{thinhnote}

By deriving these novel constraint representations, our approach enhances the theoretical analysis of constrained FB-FL control while maintaining computational tractability. The effectiveness of our framework is validated through simulation tests over a planar unmanned aerial vehicle model, stabilization of the longitudinal dynamics of aircraft and of a permanent magnet
synchronous motor, with Matlab code available at 
\url{https://gitlab.com/huuthinh.do0421/constrained-control-of-flat-systems-with-relu-ann}.

The remainder of the paper is structured as follows. Section \ref{sec:preli} specifies the context of constrained control of differentially flat systems and the groundwork to use ReLU-ANN for constraint approximation. Section \ref{sec:ConstrReform} presents the constraint characterization with ReLU-ANN and how the PWA description of the new constraints can be derived.
The integration into CLF-based and MPC of such set description will be elaborated in Section \ref{sec:control}. Simulation tests will be provided in Section \ref{sec:sim} to validate and identify the shortcoming of the proposed setting. Finally, Section \ref{sec:concl} concludes and discusses future directions.

\textit{Notation:} Bold lowercase letters denote vectors. For $\boldsymbol{x}\in\mathbb{R}^n$, ${\bx}_i$ denotes its $i$th entry.
The absolute operator is applied elementwise, $|\bx|= [|\bx_1|, \cdots, |\bx_n|]^\top$.
The usual norms are defined as $\|\bx\| = \sqrt{\bx^\top\bx}$, $\|\bx\|_1 = \sum_{i=1}^n|x_i|$. For $\rho>0$, $\mc B(\bx_0,\rho)\triangleq\{\bx:\|\bx_0-\bx\|\leq \rho\}$. \thinh{For a vector $\btt$, we denote its length (the number of entries) as $\mathrm{ len}(\btt)$.}
For a neural network, the output, bias and weight of the $\ell$th layer are denoted with the right superscript as $\boldsymbol{y}^{(\ell)}$, $\boldsymbol{b}^{(\ell)}$ and $\boldsymbol{W}^{(\ell)}$, respectively.
Bold capital letters denote matrices with appropriate dimension. For a matrix $\boldsymbol{W} $, $\boldsymbol{W}_{[i,j]}$ denotes its entry at row $i$ column $j$; $\boldsymbol{W}_{[r,:]}$ denotes its $r$-th row.
For a set $\mc X$, $|\mc X| $ denotes the its cardinality.
The ReLU activation function is denoted as $\sigma(s)=\max\{0,s\}$ and applied element-wise for a vector $\boldsymbol{x}$. A function $\nu: \R_{\geq 0}\to \R_{\geq 0}$ is of class $\mc K$ if it is continuous, strictly increasing and  positive definite. $\bQ\succ 0 \, (\bQ\succeq 0)$ signifies that $\bQ$ is positive (semi)definite. Likewise, $\bQ \prec 0 \, (\bQ \preceq 0) \Leftrightarrow -\bQ\succ 0 \, (-\bQ\succeq 0). $ 
The letter $t$ {denotes the time variable}. For some integer $k$ and a signal $\bz$, its value at time $t=kT_s$ is denoted as $\bz(k)$, for sampling time $T_s.$
Additionally, we use $\bz(k|i)$ to denote the estimation of $\bz(k+i)$ based on information available at moment $k$. \thinh{The binomial coefficient is denoted by $\begin{pmatrix}
    n \\  k
\end{pmatrix} = \dfrac{n!}{k!(n-k)!}$.}

\section{Preliminaries and problem statement}
\label{sec:preli}

The section presents the challenges of implementing constrained feedback linearization through flatness, {specifically the emergence of complex constraints due to the ancillary nonlinear coordinate changes.} Since we choose ReLU-ANN to approximate these constraints, we also provide relevant details about their structure.

\subsection{Constrained flatness-based feedback linearization}
\label{subsec:FL_formu}
\begin{figure}[htbp]
    \centering
    \begin{tikzpicture}
    \draw (0,0) node {\includegraphics[scale=1]{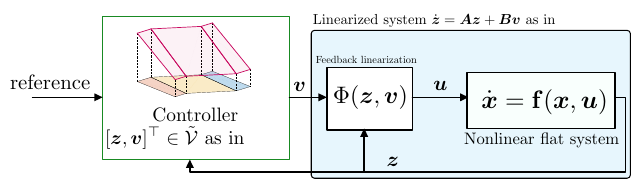}};

    \draw (-0.835,-0.85) node {\eqref{eq:Vtilde_Hspace}};

    \draw (1.79,0.49) node [scale=0.53]{\eqref{eq:flat_map}};
  \draw (4.2,1.185) node [scale=0.8]{\eqref{eq:linearized_flat}};
    
\end{tikzpicture}
    \caption{Constrained feedback linearization control scheme based on ReLU-ANN.}
    \label{fig:controlSchemePWA}
\end{figure}
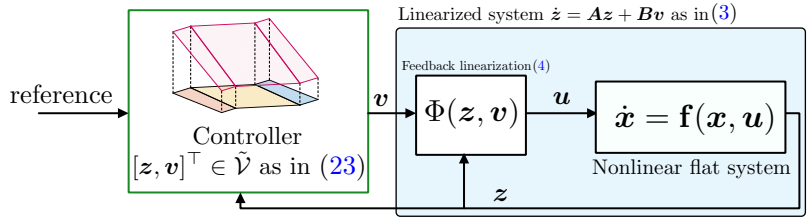
In this work, we consider the class of differentially flat systems, described by the nonlinear dynamics:
\be
\dot \bx = \bdf(\bx, \bu),
\label{eq:sys_nonlinear_orgin}
\ee 
where $\bx\in\R^n,\bu\in\R^m$ denote the state and input vectors and $\bu$ is constrained as:
\be 
\bu \in\mathcal{U}=\{\bu  \in\R^m : |\bu |\leq \bar \bu\}.
\label{eq:constr_init}
\ee 
Differentially flat (or flat) systems are well-suited for nonlinear control design as well as for trajectory planning. In this work, we focus on the property of feedback linearizability of this class \cite{levine2009analysis}. Indeed, by definition, differentially flat systems are endogenous dynamic feedback linearizable \cite[chap. 5]{levine2009analysis}. Namely, there exists a coordinate change and an input transformation such that the dynamics \eqref{eq:sys_nonlinear_orgin} can be exactly linearized into\footnote{For more details on differentially flat systems we refer to the comprehensive textbook \cite{levine2009analysis}.}:
\be 
\dot \bz = \bA \bz + \bB \bv,
\label{eq:linearized_flat}
\ee 
where $\bz\in\R^{n_z},\bv\in\R^m$ are the new state and input vector in the new coordinate system. $\bA,\bB$ are the constant matrices representing the dynamics of $m$ chains of integrators \cite{levine2009analysis} and 
\be 
\bu = \Phi(\bz,\bv) \label{eq:flat_map},
\ee 
denotes the linearizing input transformation. The feedback linearization scheme is summarized in the blue block in Fig. \ref{fig:controlSchemePWA}.
We further assume that the state vector $\bz$ can be measured, and the stabilization of \eqref{eq:linearized_flat} is implied by that of \eqref{eq:sys_nonlinear_orgin}. Then, the control problem now becomes closing the loop for the linear system \eqref{eq:linearized_flat} via state feedback $\bz$. Contrary to initial hopes, the task of stabilizing \eqref{eq:linearized_flat} is not trivial due to the constraint set
\eqref{eq:constr_init} being distorted by the variable change  induced by \eqref{eq:flat_map}. More precisely, the constraint $\bu\in \mc U$ as in \eqref{eq:constr_init} now becomes:
\be 
[\bz \ \bv]^\top \in \mc V\triangleq\left\{[\bz \ \bv]^\top :  |\Phi(\bz,\bv)|\leq \bar \bu 
\right\}, \label{eq:convoluted_constr}
\ee 
which is a joint constraint set of $(\bz$,$\bv)$ and typically difficult to characterize due to the nonlinear mapping $\Phi(\bz,\bv)$ from \eqref{eq:flat_map}.

\textit{Objective:} Within the presented setting, our goal is to synthesize an outer controller (the green block in Fig. \ref{fig:controlSchemePWA}) for the linear system \eqref{eq:linearized_flat}, subject to the convoluted constraint \eqref{eq:convoluted_constr}.

In the next section, we will show that \eqref{eq:convoluted_constr} can be decomposed into a union of polyhedra via a ReLU-ANN approximation of the mapping $\Phi(\bz,\bv)$ in \eqref{eq:flat_map}. To prepare for such constraint characterization, let us recall first the structure of the ReLU-ANN.



\subsection{Rectified linear artificial neural networks}
Consider the fully-connected ReLU-ANN with one hidden layer mapping the input $\by^{(0)}\in \mathbb{R}^{n_0} $ to the output $\by^{(2)}\in\mathbb{R}^{n_2} $ as:
\begin{equation}
         \boldsymbol{y}^{(1)} = \sigma\left(W^{(1)} \boldsymbol{y}^{(0)} + \boldsymbol{b}^{(1)}\right),\quad 
         \boldsymbol y^{{(2)}}=W^{(2)}\boldsymbol{y}^{(1)} + \boldsymbol{b}^{\thinh{(2)}},
     \label{eq:ReLU_single}
\end{equation}
where $W^{(\ell)}\in\mathbb R^{n_{\ell}\times n_{\ell-1}},\boldsymbol{b}^{(\ell)}\in\mathbb R^{ n_\ell}, \ell\in\{1,2\}$ denote the weight and bias of the network within the corresponding layers, respectively. $n_\ell$ is the number of neurons in the $\ell$th layer,
$\sigma(s)\triangleq\max(s,0), s\in\mathbb R$ and for $ \boldsymbol{y}\in\mathbb R^{n_1}$:
\begin{equation}
 \sigma(\boldsymbol{y})\triangleq[\sigma({y}_1),...,\sigma({y}_{n_1})]^\top .
\end{equation}

\begin{figure}[htbp]
    \centering
\includegraphics[width=0.7\textwidth]{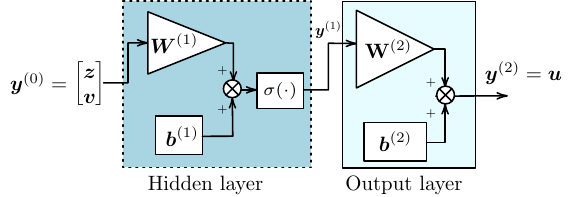}
    \caption{Single hidden layer ReLU-ANN approximating the linearizing mapping.}
    \label{fig:SingleLayer}
\end{figure}

With an approximation function based on a network of this form, in the next section, the constraint set $\mc V$ from \eqref{eq:convoluted_constr} will be partitioned into a union of non-overlapping polyhedra verifying the facet-to-facet property (a polyhedral complex \cite{jordan2019provable}). This representation will prove useful later in control synthesis.

\section{Constraint reformulation with ReLU-ANN}
\label{sec:ConstrReform}
As is well known in the literature (see \cite{hu2020analysis,samanipour2023stability,goujon2024number} for example), ReLU-ANN
can be regarded as a piecewise-affine (PWA) function associated with polyhedral disjunctive domains. For simplicity, these domains will be referred to as \textit{cells}. In the following, we show that for single-hidden-layer network as in Fig. \ref{fig:SingleLayer}, these cells can be explicitly enumerated, allowing to constructively represent convoluted constraints.

\subsection{Enumeration of cells formed by ReLU-ANN}
To pave the way to the explicit form of the PWA representation of the ReLU-ANN, we first analyze the case of a single neuron as with the following proposition.
\begin{proposition}\label{prop:oneNeuron}
    For a ReLU activation function $\sigma(s)$, denote its activation status with a binary variable $\alpha$: $\alpha=+1$ when the function is activated (i.e., $\sigma(s) = s$) and $\alpha=-1$ when the function is not activated (i.e., $\sigma(s) = 0$). 
    
    Then, the output of a neuron $y = \sigma(\boldsymbol{w}^\top\bx+b)\rule{0pt}{9pt}$ and the corresponding activated half-space can be calculated with respect to $\alpha\in\{\pm 1\}$ as:
    \begin{subnumcases}
{
\label{eq:affine_alpha}
}
    y=\frac{1}{2}(\alpha+1)(\boldsymbol{w}^\top\bx+b) &\label{eq:affine_alpha_a}\\
            \bx\in \{\bx:-\alpha \boldsymbol{w}^\top\bx \leq \alpha b\}.& \label{eq:affine_alpha_b}
\end{subnumcases}
\end{proposition}
\textit{Proof:} By replacing either $\alpha=+1$ when $\boldsymbol w^\top \bx + b\geq 0 $ or $\alpha=-1$ otherwise, we retrieve the result of the function $\sigma(\boldsymbol{w}^\top\bx+b)$.\QEDA

\begin{figure}[htb]
    \centering
    \includegraphics[width=0.625\linewidth]{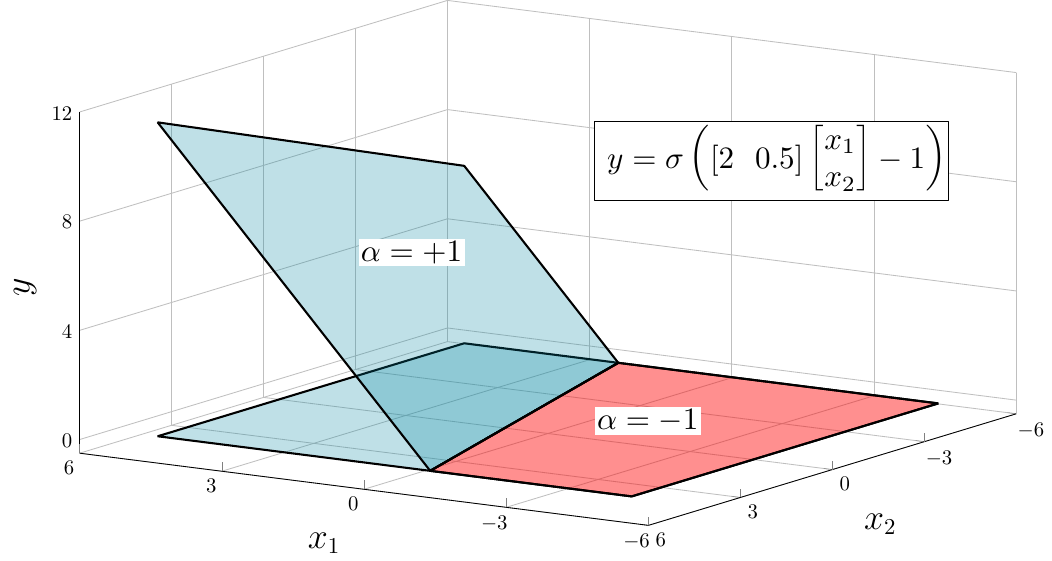}
    \caption{Illustrative example of Proposition \ref{prop:oneNeuron} with a neuron of $\bw=\bbm 2& 0.5 \ebm^\top, b=-1$.}
    \label{fig:OneNeural}
\end{figure}
With Proposition \ref{prop:oneNeuron}, one can see that the variable space of $\bx$ is partitioned into two regions as in \eqref{eq:affine_alpha_b} associated with $\alpha=-1$ and $\alpha=1$.
An illustrative example in $\R^2$ is presented in Fig. \ref{fig:OneNeural}.
This representation is relevant when one needs to enumerate the cells partitioned by the ReLU-ANN function. Namely, given a fixed value of $\alpha$, we collect an affine map associated with a half-space with \eqref{eq:affine_alpha}. Like this, the feedforward formula $y = \sigma(\boldsymbol{w}^\top\bx+b)$ can be translated to a lookup table with $\alpha\in\{\pm 1\}$. In the same manner, it will be shown that given a state of activation for all neurons in the network \eqref{eq:ReLU_single}, we can collect: 
\begin{enumerate}[i)]
    \item the affine map representing the input-output relationship of the network;
    \item the polyhedron associated with such a linear map (called the support polytope).
\end{enumerate}
More specifically, the explicit form of the PWA mappings and the associated support polyhedra are given in the following proposition.

\begin{proposition}\label{prop:explicitReLU}
Consider the activation status of the hidden layer of the network \eqref{eq:ReLU_single} and denote it as $\boldsymbol{\alpha}=(\alpha_1,\alpha_2,...,\alpha_{n_1})\in\{\pm 1\}^{n_1}$. Namely, for the $k$-th row of $\by^{(1)}$ in \eqref{eq:ReLU_single}, denoted as $\by^{(1)}_k$, $\alpha_k$ represents the activation status of the function $\by^{(1)}_k=\sigma\left(\boldsymbol{W}^{(1)}_{[k,:]}\by^{(0)}+\bb^{(1)}_k\right)$. The output $\by^{(2)}$ can then be expressed as a function of the input $\by^{(0)}$ as follows:
\be
\by^2=
\underbrace{
\bbm 
\boldsymbol F_1(\boldsymbol\alpha,\boldsymbol{W}^{(1)},\boldsymbol{W}^{(2)}) \\ \boldsymbol F_2(\boldsymbol\alpha,\boldsymbol{W}^{(1)},\boldsymbol{W}^{(2)})
\\
...
\\
\boldsymbol F_{n_2}(\boldsymbol\alpha,\boldsymbol{W}^{(1)},\boldsymbol{W}^{(2)})
\ebm }_{\mathcal F(\boldsymbol\alpha,\boldsymbol{W}^{(1)},\boldsymbol{W}^{(2)})}\by^{(0)} + 
\underbrace{\bbm 
f_1(\boldsymbol\alpha,\boldsymbol{W}^{(2)},\bb^{(1)},\bb^{(2)}) \\f_2(\boldsymbol\alpha,\boldsymbol{W}^{(2)},\bb^{(1)},\bb^{(2)})
\\
...
\\
f_{n_2}(\boldsymbol\alpha,\boldsymbol{W}^{(2)},\bb^{(1)},\bb^{(2)})
\ebm}_{\boldsymbol{f}(\boldsymbol\alpha,\boldsymbol{W}^{(2)},\bb^{(1)},\bb^{(2)})}.
\label{eq:affine_map_alp}
\ee 
where
\be
\begin{cases}
\boldsymbol F_j(\boldsymbol\alpha,\boldsymbol{W}^{(1)},\boldsymbol{W}^{(2)}) &= \sum_{k=1}^{n_1}\frac{1}{2}\boldsymbol{W}^{(2)}_{[j,k]}(\alpha_k+1)\boldsymbol{W}^{(1)}_{[k,:]}, \\
f_j(\boldsymbol\alpha,\boldsymbol{W}^{(2)},\bb^{(1)},\bb^{(2)})&=\left(\sum_{k=1}^{n_1}\frac{1}{2}\boldsymbol{W}^{(2)}_{[j,k]}(\alpha_k+1)\bb^{(1)}_k \right)+ \bb^{(2)}_j,j\in\{1,...,n_2\} .\rule{0pt}{15pt}
\end{cases}
\ee 
Moreover, the polyhedral domain associated with the mapping \eqref{eq:affine_map_alp} is described as:
\be 
\mathcal{H}(\boldsymbol\alpha,\boldsymbol{W}^{(1)},\bb^{(1)}) = \left\{\by^{(0)}:(-
\alpha_k \boldsymbol{W}^{(1)}_{[k,:]})\by^{(0)} \leq \alpha_k\boldsymbol{b}^{(1)}_k, k\in\{1,...,n_1\}
\right\}.
\label{eq:support_reg}
\ee 
\end{proposition}

\begin{proof}
    Consider the hidden layer in \eqref{eq:ReLU_single} where one has for $k\in \{1,...,n_1\}$:
    \be 
\by^{(1)}_k = \sigma\left(\boldsymbol{W}^{(1)}_{[k,:]}\by^{(0)}+\bb^{(1)}_k\right).
\label{eq:FirstLay}
    \ee
On one hand, with the activation of this node codified by $\alpha_k\in\{\pm 1\}$ as in \eqref{eq:affine_alpha}, \eqref{eq:FirstLay} can be equivalently represented by a PWA function in $\R^{n_0}$ as:
\be 
\begin{cases}
    \by_k^{(1)} = \frac{1}{2}(\alpha_k+1)(\boldsymbol{W}^{(1)}_{[k,:]}\by^{(0)}+\bb^{(1)}_k), & \\
    \by^{(0)} \in \{\by^{(0)}\in \R^{n_0}:-\alpha_k\boldsymbol{W}^{(1)}_{[k,:]}\by^{(0)} \leq \alpha_k\bb^{(1)}_k\}, \alpha_k\in \{\pm 1\}, k\in\{1,...,n_1\}.&\rule{0pt}{15pt}
\end{cases}
\label{eq:first_lay_pf}
\ee 
On the other hand, for the $j$-th entry of $\by^{(2)}$ in \eqref{eq:ReLU_single}, denoted as $\by_j^{(2)}, j \in \{1,...,n_2\}$ one has:
\be 
\by_j^{(2)}=\boldsymbol{W}^{(2)}_{[j,:]}\by^{(1)} + \bb_j^{(2)} = \sum_{k=1}^{n_1}\boldsymbol{W}^{(2)}_{[j,k]}\by_k^{(1)}+ \bb_j^{(2)}.
\label{eq:sec_lay_pf}
\ee 
Then, by replacing \eqref{eq:first_lay_pf} into \eqref{eq:sec_lay_pf}, one arrives at \eqref{eq:affine_map_alp}--\eqref{eq:support_reg}, hence completing the proof.
\end{proof}


\begin{algorithm}[htbp]
\caption{Construct the PWA representation of the ReLU-ANN}\label{algo:cellEnum}
\begin{algorithmic}[1]
\Require ReLU-ANN parameters $\bW^{(i)},\bb^{(i)}, i\in\{1,2\}$, workspace $\mc Z$.
\Ensure The list of PWA maps $\mc M$, support polytopes $\mc S$ and the corresponding binary combination $\mc A$.
\State $\mc S = [~]
  ; \mc M = [~]; \mc A = [~]
  ;$\Comment{Initialize empty lists}
\For{$\boldsymbol{\alpha} \in \{\pm 1\}^{n_1}$}
        \State Define the polyhedron $\mc H(\balpha,\bW^{(1)},\bb^{(1)})$ as in \eqref{eq:support_reg};
        \If{$\mc H(\balpha,\bW^{(1)},\bb^{(1)}) \bigcap \mc Z$ is \textbf{not} empty}
        \State Append $\mc H(\balpha,\bW^{(1)},\bb^{(1)}) \bigcap \mc Z$ into $\mc S$;\Comment{Collect the support cell}
        \State Compute $\{\mathcal F(\boldsymbol\alpha,\boldsymbol{W}^{(1)},\boldsymbol{W}^{(2)})    ,   \boldsymbol{f}(\boldsymbol\alpha,\boldsymbol{W}^{(2)},\bb^{(1)},\bb^{(2)})\}$ as in \eqref{eq:affine_map_alp};
        \State Append $\{\mathcal F(\boldsymbol\alpha,\boldsymbol{W}^{(1)},\boldsymbol{W}^{(2)})    ,   \boldsymbol{f}(\boldsymbol\alpha,\boldsymbol{W}^{(2)},\bb^{(1)},\bb^{(2)})\}$ into $\mc M$;
        \State Append $\balpha$ into $\mc A$;
\EndIf
\EndFor;

\end{algorithmic}
\end{algorithm}

\begin{figure}[htbp]
    \centering
    \includegraphics[width=0.8\linewidth]{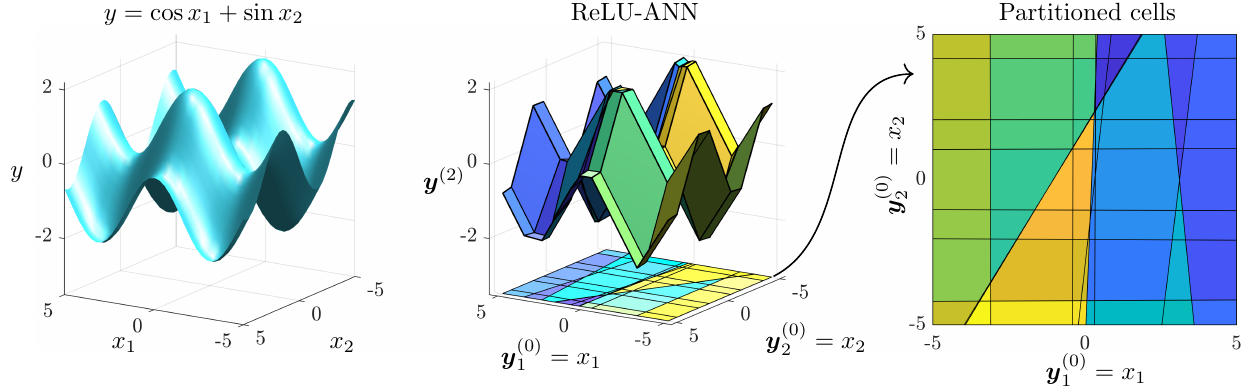}
    \caption{PWA representation of ReLU-ANN with the approximation of $y=\cos x_1 + \sin x_2.$}
    \label{fig:testMap}
\end{figure}

For a generic network of the form \eqref{eq:ReLU_single}, the procedure for cell enumeration is summarized in Algorithm \ref{algo:cellEnum}. 
It is important to point out that not all combinations of $\balpha$ lead to a non-empty cell. For this reason, a condition of non-emptiness is added at step 4 in Algorithm \ref{algo:cellEnum}. Aligned with the algorithm, $\mc A$ collects the set of $\balpha$ resulting in non-empty polyhedron cells, i.e., for $\mc H(\cdot)$ in \eqref{eq:support_reg}:
\be 
\mc A = \left\{\balpha \in \{\pm 1\}^{n_1}: \left| \mc H(\balpha,\bW^{(1)},\bb^{(1)}) \bigcap \mc Z \right|>0 \right\}, \label{eq:non-empty-index}
\ee 
where $ \mc Z$ is the user-defined set of interest in the space of the network's input $ \by^{(0)}$.
To illustrate the effectiveness of the algorithm, we approximate the function $ y = \cos x_1 +  \sin x_2 $ with a network of the form \eqref{eq:ReLU_single} and $n_1 = 15$ neurons in the hidden layer. The domain of interest $\mc Z$ is chosen as $\mc Z = \{\by^{(0)} \in \R^2: |\by^{(0)}|\leq 5 \}$.
The partitioned space and the PWA mappings are depicted in Fig. \ref{fig:testMap}.

\begin{thinhnote}
    It can be noted that the number of cells derived or the complexity of Algorithm \ref{algo:cellEnum} increases significantly with the size of the hidden layer. This can be noted via the lower bound on the maximal number of cells
\cite{karg2020efficient,montufar2014number}:
\be
\left(\prod_{l=1}^{L-1}\left\lfloor\frac{M}{n_x}\right\rfloor^{n_x}\right) \sum_{j=0}^{n_x}\binom{L}{j}, \label{eq:BoundBoundNN}
\ee 
where $L $ is the number of hidden layers and $M$ is the number of neurons in each layer.
    
\end{thinhnote}

\begin{remark}
\label{rem:NN_PWA}

\thinh{
While the numerical enumeration of PWA maps and their cells from a ReLU-ANN is not new in the literature, the rationale behind Proposition \ref{prop:explicitReLU}  is to extract these PWA functions explicitly, which is essential for the subsequent construction. This explicit formulation makes the network representation transparent and facilitates the developments that follow.}
    \end{remark}

\begin{figure}[htbp]
    \centering
    \resizebox{0.9\textwidth}{!}{
\begin{tikzpicture}
    \draw (0,0) node {\includegraphics[scale=1]{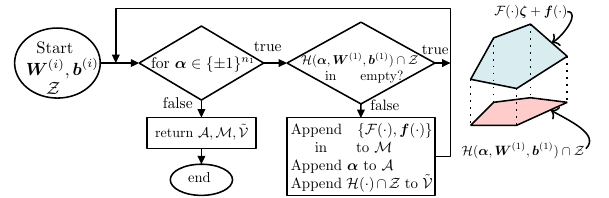}};

    \draw (0.65,-0.85) node [scale=0.65]{\eqref{eq:affine_map_alp}};

    \draw (0.75 ,0.365) node [scale=0.6]{\eqref{eq:support_reg}};
\end{tikzpicture}
    }
    \caption{Cell enumeration algorithm (Algorithm \ref{algo:cellEnum}) applied for the approximation in \eqref{eq:tildeV_zv}.}
    \label{fig:algoCell}
\end{figure}

So far, we have presented a procedure to compute the PWA representation of a ReLU-ANN. {In the next section, we will place this groundwork within} a control-theoretic context to {address the constrained control problem originally defined in} Section \ref{subsec:FL_formu}.

\subsection{Constraint characterization}
We now show that the convoluted constraint set $\mc V$ can be inner approximated by a union of polyhedra, which are deduced from the PWA representation of a ReLU-ANN. This constraint processing gives rise to different optimization-based control designs, including explicit and implicit MPC.



First, recall that the constraint \eqref{eq:convoluted_constr} can be implied by bounding the output of the ReLU-ANN approximation
of $\Phi(\bz,\bv)$ from \eqref{eq:flat_map}.

\begin{proposition}[Convoluted constraint enforcement with ReLU-ANN \cite{do2024_reluANN}] \label{prop:constraint_convoluted}

    Consider a ReLU-ANN approximation of $\Phi(\bz,\bv)$ in \eqref{eq:flat_map} with the structure \eqref{eq:ReLU_single}, called $\Phi_{nn}(\bz,\bv)$ satisfying:
    \be 
|\Phi(\bz,\bv) - \Phi_{nn}(\bz,\bv)|\leq \bep. \label{eq:assum_error}
    \ee 
Then, the constraint \eqref{eq:convoluted_constr} is implied by the tightened form:
\be 
|\Phi_{nn}(\bz,\bv)| \leq \bar \bu  - \bep.
\label{eq:MI_constr}
\ee 
In other words,
\be 
\tilde{ \mc V} = \{[\bz\ \bv]^\top:|\Phi_{nn}(\bz,\bv)| \leq \bar \bu  - \bep\} \subset \mc V. \label{eq:tildeV_ANN}
\ee 
\end{proposition}

\begin{proof}
   The proof can be simply deduced by showing that the condition \eqref{eq:MI_constr} implies the satisfaction of \eqref{eq:convoluted_constr}. First, let us use the subscript $i$ to denote the $i$th row of a vector. Then, we have:
   \be 
   \begin{aligned}
       &|\bu_i|=|\Phi_i(\bz,\bv)| = |\Phi_i(\bz,\bv)-\Phi_{nn,i}(\bz,\bv)+\Phi_{nn,i}(\bz,\bv)|  \\
       & \leq |\Phi_i(\bz,\bv)-\Phi_{nn,i}(\bz,\bv)|+|\Phi_{nn,i}(\bz,\bv)| \leq \bep_i +|\Phi_{nn,i}(\bz,\bv)|.
   \end{aligned}
   \ee 
   Then, in the vector form we have:
   \be 
|\bu |\leq |\Phi_{nn,i}(\bz,\bv)|+ \bep.
   \ee 
   Consequently, one can state that \eqref{eq:MI_constr} is a sufficient condition for \eqref{eq:convoluted_constr}, or:
   \be 
[\bz \ \bv]^\top \in \tilde{\mc V} \text{ in \eqref{eq:tildeV_ANN}} \Rightarrow [\bz \ \bv]^\top \in {\mc V} \text{ in \eqref{eq:convoluted_constr} }\Leftrightarrow \bu \in \mc U \text{ in \eqref{eq:constr_init}}.
   \ee 
\end{proof}

\begin{thinhnote}
    In this work, the condition \eqref{eq:assum_error} is adopted as a standing assumption, consistent with common practice in learning-based control (see, e.g., \cite{fabiani2022reliably,ge2007approximation}). 
    The error can be validated in practice empirically via sampling-based \cite{sivaramakrishnan2025saver} or formally via interval arithmetic \cite{ehlers2017formal,li2025control}.
The verification of neural networks remains an active research topic, and establishing rigorous bounds is still challenging (see the recent survey \cite{rossig2021advances} for further discussion).
    
\end{thinhnote}

\begin{remark}
In \cite{do2024_reluANN}, it was shown that the constraint
  \eqref{eq:MI_constr} can be rewritten into a set of mixed-integer (MI) linear constraints under the assumption that the upper and lower bounds of each neurons' output can be found. In this work, as shown in the following, such an assumption will be deliberately avoided by constructing the explicit expression of the partitioned polyhedron cells. 
\end{remark}
Hereinafter, let us denote $\bzeta\triangleq [\bz \ \bv]^\top$, then with Proposition \ref{prop:constraint_convoluted}, the problem of imposing 
$\bzeta\in\mc V$ as in \eqref{eq:convoluted_constr} is translated to $\bzeta \in \tilde{ \mc V}$ as in \eqref{eq:tildeV_ANN}.
Meanwhile, with the space partitioned by $\Phi_{nn}(\bz,\bv)$ and the corresponding cells collected with Proposition \ref{prop:explicitReLU}, one can obtain an explicit representation of $\tilde{\mc V} $ in \eqref{eq:tildeV_ANN} as the following:
\be 
\tilde{\mc V} =\bigcup_{\balpha\in\{\pm 1\}^{n_1} }\hspace{-0.3cm}\left\{ \bzeta : \left| \mc F(\boldsymbol\alpha,\boldsymbol{W}^{(1)},\boldsymbol{W}^{(2)}) \bzeta + \bff (\balpha,\bW^{(2)},\bb^{(1)},\bb^{(2)})\right|
\leq \bar \bu - \bep
\right\}, \label{eq:tildeV_zv}
\ee 
with $\mc F(\cdot), \bff(\cdot)$ from \eqref{eq:affine_map_alp}.

In this way, the set $\tilde{\mc V} $ can be represented by a union of polyhedra.
Without loss of generality, we rewrite $\tilde{\mc V}$ as a union of $|\mc A|$ polyhedrons, where the set $\mc A$ contains the combination of $\balpha$ creating nonempty cells as in \eqref{eq:non-empty-index}, and $|\mc A|$ denotes its number of elements. Namely, we denote $|\mc A|$ cells of $\tilde{\mc V}$ in \eqref{eq:tildeV_zv} as following:
\be 
\tilde{\mc V} = \bigcup_{j=1}^{|\mc A|}\mc C_j \triangleq \bigcup_{j=1}^{|\mc A|}\left\{\bzeta=\bbm \bz \\ \bv\ebm :\Theta_j\bzeta \leq \btt_j\right\},
\label{eq:Vtilde_Hspace}
\ee 
where $\Theta_j$ and $\btt_j$ are the numerical values describing the non-empty cell in \eqref{eq:tildeV_zv}.
Then, the set-based constraints in \eqref{eq:Vtilde_Hspace} can be 
translated to MIP, by employing the new integer variables\footnote{or boolean variables to be precise.} $\beta_j\in\{0;1\}$ as \cite{prodan2015mixed}:

   \begin{subnumcases}
{
\label{eq:MIPrep}
}
     \Theta_j\bzeta \leq \btt_j + \beta_jM_j ,& \label{eq:bigM_general}\\
   \beta_j \in \{0,1\}, j\in \{1,...,|\mc A|\}
   &\\
   \sum_{j=1}^{|\mc A|}\beta_j  = |\mc A|-1,\label{eq:oneActive}&
\end{subnumcases}
where $M_j$ are the so-called ``big-M'' parameters, or sufficiently large\footnote{\thinh{The selection of the big-M constant to ensure such property is a well-discussed topic in the literature. We refer to the Appendix \ref{app:BigM} for the verification procedure.}} constants such that when $\beta_j=1$, the constraint \eqref{eq:bigM_general} becomes redundant. 
In this way, by assigning each region $\mc C_j$ with a binary variable $\beta_j$ and enforcing condition \eqref{eq:oneActive}, one ensures that there will be {only one active constraint} in \eqref{eq:bigM_general}, or equivalently, $\bzeta\in\tilde{\mc V}$ as in \eqref{eq:Vtilde_Hspace}.


The representation \eqref{eq:MIPrep}, will be later employed to synthesize the controllers for the system. Yet, before entering into the control design, let us analyze first an important technical concern of Proposition \ref{prop:constraint_convoluted} stemming from the supposition \eqref{eq:assum_error}.




\subsection{Discussion on error bound estimation approaches}
\label{subsec:err_bound}
Clearly, the inclusion $\tilde{\mc V} \subset \mc V$ in \eqref{eq:tildeV_ANN} hinges on the guarantee of approximation error $\bep$ in \eqref{eq:assum_error} for a compact domain of interest $(\bz,\bv)\in \mc Z \subset \R^{n_z}\times \R^m $. 
In the literature, it is notable that the error bound estimation remains open \cite{elbrachter2021deep,liang2022deep,franco2023approximation}.
In this work, under the assumption that the neural network has been a priori constructed, some analysis on the possible estimation of $\bep $ can be given as the following.

First, the most straightforward attempt is to solve the optimization:
\be 
\bep_i=\max_{[\bz,\bv]^\top\in\mc Z} |\Phi_{i}(\bz,\bv) - \Phi_{nn,i}(\bz,\bv)|, \label{eq:opti_bep}
\ee 
where the subscript $i$ refers to the $i$th row of the vector.
In general, \eqref{eq:opti_bep} is not trivial to solve, since this is possibly a non-convex optimization problem due to the typical complexity of $\Phi(\cdot)$ in feedback linearization-based control. Furthermore, with the piecewise affine nature of $\Phi_{nn}(\cdot)$, the cost function of \eqref{eq:opti_bep} is certainly non-smooth, hindering the application of gradient-based algorithms.  This direction may prove efficient for small scale system with gradient-free optimization based approaches.

Another approach, based on brute-force search for the maximizer of \eqref{eq:opti_bep}, is to discretize the search space into a grid of sample points. Then, the optimal solution can be estimated
by checking the objective function 
over the grid with assumptions on the Lipschitz constant characterizing the nonlinear function. \thinh{In other words, as a standard technique in the literature, one can compute an upper bound of the error for all points in the set with the following proposition.}

\begin{proposition}[Grid-based error estimation]
\label{prop:gridbasedError}

Denote the vector $\bzeta \triangleq \bbm \bz & \bv \ebm^\top$ and the approximation error $\bvep(\bzeta) \in \R^m$:
\be 
\bvep(\bzeta) = \Phi(\bz,\bv)-\Phi_{nn}(\bz,\bv). \label{eq:error_func}
\ee 
Suppose that the $i$th row of $\bvep(\bzeta)$, ($1\leq i \leq m$) is Lipschitz continuous, i.e., $\exists \gamma_i >0: $ 
\be 
| \bvep_i(\bzeta_a) - \bvep_i(\bzeta_b)| \leq \gamma_i \|\bzeta_a-\bzeta_b\|. \label{eq:lipschitz_approx_err}
\ee 
Then, given a set of grid (sample) points $\mc Z_g\subset \mc Z$ satisfying:
\be 
|\bvep_i(\bzeta_g) | \leq \tilde\bep_{i},~ \forall \bzeta_g \in \mc Z_g, \label{eq:gridBounds}
\ee 
an upper bound of $\bvep_i(\bzeta)$ over $\mc Z$ can be deduced as:
\be 
|\bvep_i(\bzeta) |\leq \tilde\bep_{i}+{\gamma_i \bar \rho },~\forall\bzeta\in \mc Z,\label{eq:estimation_grid}
\ee 
where $\bar\rho = \sup_{\bzeta\in \mc Z}\rho(\bzeta)$ and $\rho(\bzeta)\triangleq\inf_{\bzeta_g\in\mc Z_g}\|\bzeta - \bzeta_g\|$ is the granularity function defined over the grid $\mc Z_g$. 
\end{proposition}
\begin{proof}
    Consider $\bzeta\in \mc Z, \bzeta_g \in \mc B(\bzeta, \bar \rho)\cap \mc Z_g$, the $i$th row of the error \eqref{eq:error_func} can be bounded as:
    \be \begin{aligned}
        & |\bvep_i(\bzeta)| = |\bvep_i(\bzeta) -\bvep_i(\bzeta_g)+ \bvep_i(\bzeta_g)| \leq |\bvep_i(\bzeta) -\bvep_i(\bzeta_g)|+ |\bvep_i(\bzeta_g)| \\ 
        & \leq |\bvep_i(\bzeta) -\bvep_i(\bzeta_g)|+ \tilde{\bep}_i \leq \gamma_i\|\bzeta -\bzeta_g\|+ \tilde{\bep}_i \leq \gamma_i\bar\rho+ \tilde{\bep}_i.
    \end{aligned}
    \ee 
    where the last inequality holds by the definition of the granularity function $\rho(\bzeta)$ over the grid {$\mc Z_g$ as in \eqref{eq:estimation_grid}}.
\end{proof}
By reducing $\bar \rho$ or increasing the density of the grid point $ \mc Z_g$, one can obtain a more accurate estimation using  \eqref{eq:estimation_grid}.
It is noteworthy that the assumption of Lipschitz continuity in \eqref{eq:lipschitz_approx_err} is relatively not restrictive and directly boils down to the Lipschitz continuity of $\Phi(\cdot)$.
This is because the Lipschitz constant of the ANN can be computed in closed form thanks to its PWA representation \cite{rockafellar2009variational,fabiani2024robust}.
Furthermore, leveraging the PWA representation of the network, the approximation error can be estimated over each cell with the next proposition.

\begin{figure}[htbp]
    \centering
\resizebox{0.70\textwidth}{!}{
\definecolor{skyblue}{HTML}{2ebedb}

\tikzset{every picture/.style={line width=0.75pt}} 

\begin{tikzpicture}[x=0.75pt,y=0.75pt,yscale=-1,xscale=1, line join = bevel]

\draw  [color={rgb, 255:red, 27; green, 33; blue, 111 }  ,draw opacity=1] (295,55) -- (360,55) -- (350,90) -- (270,90) -- cycle ;
\draw    (250,140) -- (241.41,148.59) ;
\draw [shift={(240,150)}, rotate = 315] [color={rgb, 255:red, 0; green, 0; blue, 0 }  ][line width=0.75]    (4.37,-1.32) .. controls (2.78,-0.56) and (1.32,-0.12) .. (0,0) .. controls (1.32,0.12) and (2.78,0.56) .. (4.37,1.32)   ;
\draw    (250,140) -- (250,107) ;
\draw [shift={(250,105)}, rotate = 90] [color={rgb, 255:red, 0; green, 0; blue, 0 }  ][line width=0.75]    (4.37,-1.32) .. controls (2.78,-0.56) and (1.32,-0.12) .. (0,0) .. controls (1.32,0.12) and (2.78,0.56) .. (4.37,1.32)   ;
\draw    (250,140) -- (263,140) ;
\draw [shift={(265,140)}, rotate = 180] [color={rgb, 255:red, 0; green, 0; blue, 0 }  ][line width=0.75]    (4.37,-1.32) .. controls (2.78,-0.56) and (1.32,-0.12) .. (0,0) .. controls (1.32,0.12) and (2.78,0.56) .. (4.37,1.32)   ;
\draw  [ fill = skyblue, fill opacity = 0.075 ] (360,145) -- (350,165) -- (270,165) -- (295,140) -- cycle ;
\draw  [dash pattern={on 0.84pt off 2.51pt}]  (295,55) -- (295,140) ;
\draw  [dash pattern={on 0.84pt off 2.51pt}]  (270,90) -- (270,165) ;
\draw  [dash pattern={on 0.84pt off 2.51pt}]  (350,90) -- (350,165) ;
\draw  [dash pattern={on 0.84pt off 2.51pt}]  (360,49) -- (360,145) ;
\draw  [color={rgb, 255:red, 227; green, 42; blue, 42 }  ,draw opacity=1 ][fill={rgb, 255:red, 255; green, 130; blue, 130 }  ,fill opacity=0.2 ] (295,70) .. controls (300.7,51.77) and (343.03,47.77) .. (360,60) .. controls (344.7,54.44) and (350.7,88.44) .. (350,95) .. controls (328.03,111.77) and (295.7,128.11) .. (270,95) .. controls (286.36,92.77) and (292.7,81.44) .. (295,70) -- cycle ;
\draw  [color={rgb, 255:red, 61; green, 177; blue, 107 }  ,draw opacity=1 , fill = {rgb, 255:red, 61; green, 177; blue, 107 } , fill opacity = 0.1] (295,49) -- (360,49) -- (350,104) -- (270,105) -- cycle ;
\draw [color={rgb, 255:red, 27; green, 33; blue, 111 }  ,draw opacity=1 ]   (270,90) -- (350,90) ;

\draw  [color={rgb, 255:red, 227; green, 42; blue, 42 }  ,draw opacity=1 ][fill={rgb, 255:red, 227; green, 42; blue, 42 }  ,fill opacity=0.33 ] (75,85) -- (91.17,85) -- (91.17,91.65) -- (75,91.65) -- cycle ;
\draw  [color={rgb, 255:red, 27; green, 33; blue, 111 }  ,draw opacity=1 ] (75,106.35) -- (91.17,106.35) -- (91.17,113) -- (75,113) -- cycle ;
\draw  [color={rgb, 255:red, 61; green, 177; blue, 107 }  ,draw opacity=1 ] (75.17,142.35) -- (91.34,142.35) -- (91.34,149) -- (75.17,149) -- cycle ;

\draw (96.17,81) node [anchor=north west][inner sep=0.75pt]  [scale=0.75] [align=left] {$\displaystyle \Phi_i(\boldsymbol{z} ,\boldsymbol{v})$ true function};
\draw (96.17,101) node [anchor=north west][inner sep=0.75pt]  [scale=0.75] [align=left] {$\displaystyle \Phi_{nn,i}(\boldsymbol{z} ,\boldsymbol{v})$ ReLU PWA \\approximation};
\draw (95.17,138) node [anchor=north west][inner sep=0.75pt]  [scale=0.75] [align=left] {$\displaystyle \phi (\boldsymbol{z} ,\boldsymbol{v})$ 1st order\\ approximation};
\draw (270,161.5) node  [font=\scriptsize, shift={(0,-3)}  ]  {$\bullet $};
\draw (350,162.5) node  [font=\scriptsize, shift={(0,-3)}  ]  {$\bullet $};
\draw (295,136.5) node  [font=\scriptsize, shift={(0,-3)}  ]  {$\bullet $};
\draw (360,141.5) node  [font=\scriptsize, shift={(0,-3)}  ]  {$\bullet $};
\draw (270,92.5) node  [font=\scriptsize, shift={(0,-3)}  ,color={rgb, 255:red, 227; green, 42; blue, 42 }  ,opacity=1 ]  {$\bullet $};
\draw (270,102.5) node  [font=\scriptsize, shift={(0,-3)}  ,color={rgb, 255:red, 61; green, 177; blue, 107 }  ,opacity=1 ]  {$\bullet $};
\draw (350,92) node  [font=\scriptsize, shift={(0,-3)}  ,color={rgb, 255:red, 227; green, 42; blue, 42 }  ,opacity=1 ]  {$\bullet $};
\draw (360,58) node  [font=\scriptsize, shift={(0,-3)}  ,color={rgb, 255:red, 227; green, 42; blue, 42 }  ,opacity=1 ]  {$\bullet $};
\draw (295,65.5) node  [font=\scriptsize, shift={(0,-3)}  ,color={rgb, 255:red, 227; green, 42; blue, 42 }  ,opacity=1 ]  {$\bullet $};
\draw (350,101.5) node  [font=\scriptsize, shift={(0,-3)}  ,color={rgb, 255:red, 61; green, 177; blue, 107 }  ,opacity=1 ]  {$\bullet $};
\draw (360,46) node  [font=\scriptsize, shift={(0,-3)}  ,color={rgb, 255:red, 61; green, 177; blue, 107 }  ,opacity=1 ]  {$\bullet $};
\draw (295,45.5) node  [font=\scriptsize, shift={(0,-3)}  ,color={rgb, 255:red, 61; green, 177; blue, 107 }  ,opacity=1 ]  {$\bullet $};
\draw (270,86.5) node  [font=\scriptsize, shift={(0,-3)}  ,color={rgb, 255:red, 27; green, 33; blue, 111 }  ,opacity=1 ]  {$\bullet $};
\draw (350,86.5) node  [font=\scriptsize, shift={(0,-3)}  ,color={rgb, 255:red, 27; green, 33; blue, 111 }  ,opacity=1 ]  {$\bullet $};
\draw (360,52.5) node  [font=\scriptsize, shift={(0,-3)}  ,color={rgb, 255:red, 27; green, 33; blue, 111 }  ,opacity=1 ]  {$\bullet $};
\draw (295,52) node  [font=\scriptsize, shift={(0,-3)}  ,color={rgb, 255:red, 27; green, 33; blue, 111 }  ,opacity=1 ]  {$\bullet $};
\draw (231,141.4) node [anchor=north west][inner sep=0.75pt]  [scale=0.75]  {$\boldsymbol{z}$};
\draw (256,129) node [anchor=north west][inner sep=0.75pt]  [scale=0.75]  {$\boldsymbol{v}$};

\draw (290,152) node [anchor=north west][scale=0.7][inner sep=0.75pt]    {$\mathcal{C}_j$ in \eqref{eq:Vtilde_Hspace}};

\end{tikzpicture}
}
    \caption{Illustration of Proposition \ref{prop:error_esti}.}
    \label{fig:taylor}
\end{figure}
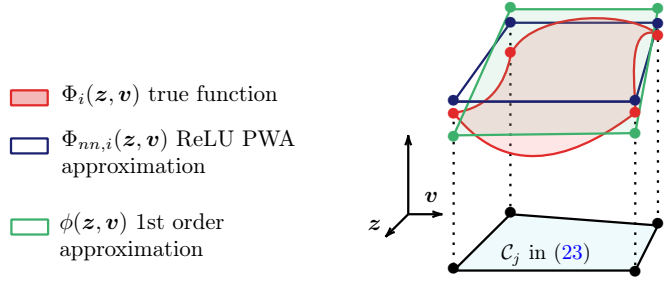

\begin{proposition}
        \label{prop:error_esti}
Inside the support polyhedron $\mc C_j$ as in \eqref{eq:Vtilde_Hspace}, the approximation error between ReLU-ANN and the real function can be upper bounded as follows:
\begin{equation}
    \sup_{[\boldsymbol{z},\boldsymbol{v}]^\top\in \mathcal{C}_j} |\Phi_i(\boldsymbol{z}, \boldsymbol{v}) - \Phi_{nn,i}(\boldsymbol{z}, \boldsymbol{v}) | \leq \Bar{\epsilon}_T+ \Bar{\epsilon}_H.
\end{equation}
where $\Bar{\epsilon}_T$ is the upper bound between the Taylor linear approximation $\phi(\bz,\bv)$ of $\Phi_i(\boldsymbol{z}, \boldsymbol{v})$ and its true value:
\begin{equation}
\Bar{\epsilon}_T = \sup_{[\boldsymbol{z},\boldsymbol{v}]^\top\in \mc C_j} |   \Phi_i(\boldsymbol{z}, \boldsymbol{v}) -  \phi(\boldsymbol{z}, \boldsymbol{v})|.
\end{equation}
$\Bar{\epsilon}_H $ is the largest approximation error evaluated at the vertices of $\mathcal{C}_j$ in \eqref{eq:Vtilde_Hspace}:
\begin{equation}
\Bar{\epsilon}_H = \sup_{[\boldsymbol{z},\boldsymbol{v}]^\top\in \mathrm{vert}(\mc C_j)} |   \Phi_{nn,i}(\boldsymbol{z}, \boldsymbol{v}) -  \phi(\boldsymbol{z}, \boldsymbol{v})|.
\label{eq:ErrorPWA_Taylor}
\end{equation}
\end{proposition}
\begin{proof}
By introducing the Taylor approximation term, one can obtain:
    \begin{equation}
    \begin{aligned}
        & |\Phi_i(\boldsymbol{z}, \boldsymbol{v}) - \Phi_{nn,i}(\boldsymbol{z}, \boldsymbol{v}) | 
        = |\Phi_i(\boldsymbol{z}, \boldsymbol{v}) -\phi(\boldsymbol{z}, \boldsymbol{v}) - (\Phi_{nn,i}(\boldsymbol{z}, \boldsymbol{v}) -\phi(\boldsymbol{z}, \boldsymbol{v}))| 
        \\
        &\leq
        |\Phi_i(\boldsymbol{z}, \boldsymbol{v}) -\phi(\boldsymbol{z}, \boldsymbol{v})| +| \Phi_{nn,i}(\boldsymbol{z}, \boldsymbol{v}) -\phi(\boldsymbol{z}, \boldsymbol{v})| 
        \leq 
        \Bar{\epsilon}_T + \Bar{\epsilon}_H.
    \end{aligned}
    \end{equation}
\end{proof}


{Estimating $\bar \epsilon_T$ is a standard problem and reduces to finding an upper bound for the Taylor's residual.} More specifically, about a linearization point $\bzeta_e$, the approximation error can be bounded as \cite{folland1990remainder}: 
    \be 
\Bar{\epsilon}_T=|\Phi_i(\bzeta) -\phi(\bzeta)| \leq \frac{C_\zeta}{2}\|\bzeta - \bzeta_e\|_1                         \label{eq:ErrorTaylorBound}      
    \ee
where the constant $C_\zeta$ is from the continuity assumption: \be 
 \left|\frac{\partial\Phi_i}{\partial\bzeta_k}(\bzeta_e + \Delta)-\frac{\partial\Phi_i}{\partial\bzeta_k}(\bzeta_e)\right|\leq C_\zeta\|\Delta\|_1, \forall k\in\{1,...,\dim(\bzeta)\}. \label{eq:SmoothAssumption}\ee 
Additionally, $\Bar{\epsilon}_H$ can be computed by simply evaluating the error $|\Phi_{nn,i}(\boldsymbol{z}, \boldsymbol{v}) -\phi(\boldsymbol{z}, \boldsymbol{v})|$ at the vertices of $\mc C_j$ in \eqref{eq:Vtilde_Hspace}, and collecting the largest value.

With the above discussion, the approximation error of the convoluted constraint with ReLU-ANN can be estimated. In the next section, we analyze the established setting within the common optimization-based frameworks of Model Predictive Control (MPC) and Control Lyapunov Function (CLF).

\section{Control implementation with ANN-based constraints}
\label{sec:control}

In the literature, it is understood that there are multiple optimization-based solutions to tackle 
the constraint \eqref{eq:Vtilde_Hspace}.
In \cite{do2024_reluANN}, for example, Control Lyapunov Functions (CLF) was used to identify the stabilizing subset of the input space $\bv$. Like this, the online optimization problem may present a MI linear or quadratic program.
Another computationally efficient solution is to regard the constraint \eqref{eq:Vtilde_Hspace} as a union of linear constraints \cite{romagnoli2020new} united by an ``OR'' operator (i.e., enforcing the state-input pair to be in one polyhedron cell).
This control strategy is based on
a line search for  an adaptive scalar gain in the interval $[0;\,1]$, {i.e., it consists in} scaling the reference signal appropriately to avoid constraint violation.
However, in this work, we would like to focus on implicit approaches where the controller is computed from an online optimization problem.

Hence, in what follows, we limit ourselves with the integration of the ANN-based constraints into two common control strategies: CLF and MPC.

\subsection{CLF implementation}

The linearized dynamics of \eqref{eq:linearized_flat} are a particular case of a control-affine system, enabling the use of a CLF-based technique. A standard setting can be adopted as \cite{li2023survey}:
\begin{subequations}
\label{eq:CLF-general}
\begin{align}
&\min_{\bv} \|\bv - \bv_d(\bz)\|^2 \\
\text{s.t }&\bbm \bz \\ \bv \ebm\in \tilde{\mc V} \text{ in \eqref{eq:Vtilde_Hspace}},  \\
&\nabla V(\bz)^\top (\bA\bz + \bB \bv) \leq -\nu(\|\bz\|),
\end{align}
\end{subequations}
where  $V(\bz)$ is a CLF of the linear system \eqref{eq:linearized_flat}, $\nu(\cdot)$ is a $\mc K$-class function and $\bv_d(\bz)$ is a desired control action we want to follow.

Given that we have linear dynamics, this solution provides a means to stabilize the system with the standard procedure of finding $V(\bz)$. For example, one can use a quadratic function:
\be 
V(\bz) = \bz^\top \bP \bz,
\label{eq:quadCLF}
\ee 
where $\bP = \Psi^{-1}$, and $\Psi$ is a solution of the linear matrix inequality \cite{blanchini2008set}:
\be 
\Psi\bA^\top + \bA \Psi  - 2\bB\bB^\top \preceq  -\gamma \Psi,\quad \Psi \succ 0,
\label{eq:LMI}
\ee 
with some $\gamma>0 $ specifying the exponential convergence rate of the closed-loop with $\bv = -\bB^\top \bP \bz$. Then, adopting the MIP representation in \eqref{eq:MIPrep}, the complete online program can be written as:
\begin{subequations}
\label{eq:CLF-general-MIP}
\begin{align}
&\min_{\bv} \|\bv - \bv_d(\bz)\|^2 \\
\text{s.t }&     \Theta_j\bzeta \leq \btt_j + \beta_jM_j ,\label{eq:CLF_input_condi1}\\
  &\beta_j \in \{0,1\}, j\in \{1,...,|\mc A|\}
   \label{eq:CLF_input_condi2}\\
   &\sum_{j=1}^{|\mc A|}\beta_j  = |\mc A|-1,  \label{eq:CLF_input_condi3}\\
&
2\bz^\top\bP (\bA\bz + \bB \bv) \leq -\gamma\bz^\top\bP\bz , \label{eq:CLF_stabi_condi}
\end{align}
\end{subequations}
{where $\bv_d(\bz)$ represents a desired high-performance control input that we aim to follow. Intuitively, the program \eqref{eq:CLF-general-MIP} is to find the closest vector $\bv$ with respect to $\bv_d(\bz)$ (i.e., a projection of $\bv_d(\bz)$) inside the set of stabilizing and admissible input described by \eqref{eq:CLF_stabi_condi} and \eqref{eq:CLF_input_condi1}--\eqref{eq:CLF_input_condi3}, respectively.}
As can be seen from the cost and constraints, the optimization problem \eqref{eq:CLF-general-MIP} presents a mixed-integer quadratic program (MIQP). While the problem is nonconvex, it admits tractable solutions through the use of appropriate optimization algorithms \cite{gurobi,quirynen2021sequential}. 
With this technique, a stabilizing condition can be integrated as in \eqref{eq:CLF_stabi_condi} while the input constraint is enforced via \eqref{eq:CLF_input_condi1}--\eqref{eq:CLF_input_condi3}.
 The only limitation of this design is its inability to handle state constraints.  Consequently, the subsequent section presents an MPC approach that explicitly incorporates these constraints over a predicted future horizon.

\subsection{MPC implementation}
\label{subsec:MPCForm}

\begin{figure}[htbp]
    \centering
    \includegraphics[width=0.5\linewidth]{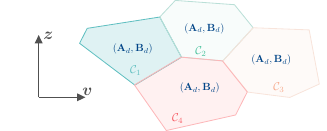}
    \caption{Interpretation of the proposed setting in the context of PWA dynamics.}
    \label{fig:PWADyna}
\end{figure}

In this part, we formulate the MPC problem within the framework of MIP.
In general, being solved in a receding horizon fashion, the online control optimization problem will be formulated as follows:
\begin{subequations}
\label{eq:impMPC_MIP}
\begin{align}
&\min_{\bv(k|\cdot)}\sum_{i=0}^{N_p-1}\|\bz(k|i) \|_{\bQ}^2 + \|\bv(k|i)\|_{\bR}^2 
\\
&
\bz(k|0) = \bz_0, \\
    &\bz(k|i+1) = \bA_d \bz(k|i) + \bB_d\bv(k|i), \ 
 \\
&\bz(k|i) \in \mc Z_s, i \in \{0,
..., N_p-1\},\\
&[\bz(k|i) ,\bv(k|i) ]^\top \in \tilde{\mc V} \text{ as in \eqref{eq:Vtilde_Hspace}},\label{eq:unionConditionMPC_imp}
\end{align}
\end{subequations}
where $\bA_d,\bB_d$ represent the discrete prediction model of \eqref{eq:linearized_flat}, $\bz_0$ is the feedback value at time step $k$ and the prediction sequence's length is noted as $N_p$. The set $\tilde{\mc V}$ is defined as in  \eqref{eq:Vtilde_Hspace}, and $\mc Z_s$ denotes the state constraints in the linearizing space. 
\begin{remark}
    With the constraint \eqref{eq:unionConditionMPC_imp}, it is implied that all the predicted values of $\Phi(\bz(k|k+i) ,\bv(k|k+i))$ or $\bu(k|k+i)$ satisfy the constraint \eqref{eq:constr_init} for all $i \in \{0,
..., N_p-1\}$, unlike the feedback linearization-based MPC settings in \cite{nevistic1994feasible,kurtz1998feedback,kandler2012differential}. This implication is relevant because ensuring constraint satisfaction for the entire predicted trajectory is a fundamental assumption to analyze closed-loop stability or recursive feasibility \cite{chen1998quasi}.

\end{remark}
Next, we show that, even when propagated through the prediction horizon, the constraint \eqref{eq:unionConditionMPC_imp} remains MI linear constraints, making the problem \eqref{eq:impMPC_MIP}, again an MIQP.
Indeed, adapted from the formulation \eqref{eq:MIPrep} the constraint \eqref{eq:unionConditionMPC_imp} can be rewritten in the form of MI linear constraints as:
\be 
\begin{cases}
   \Theta_j \bbm \bz(k|i) \\ \bv(k|i)\ebm \leq \btt_j + \beta_{ij} M_j, 
   \beta_{ij}\in \{0,1\}, j  \in\{0,...,|\mc A|\}& \\
   \sum_{j=1}^{|\mc A|}\beta_{ij} = |\mc A|-1, i \in \{0,...,N_p-1\}. \rule{0pt}{13pt}
\end{cases}
\label{eq:predictedConstraint}
\ee 
Then, by replacing \eqref{eq:predictedConstraint}, into \eqref{eq:impMPC_MIP}, we arrive to an MIQP with quadratic cost and MI linear constraints. The interpretation of \eqref{eq:predictedConstraint} is that of the predicted state and input pair $(\bz(k|i),\ \bv(k|i))$ can only stay in one cell in each prediction time step. 
Interestingly, this aligns with the concept of piecewise-affine (PWA) systems \cite{sontag1981nonlinear,Christophersen2007}, where the state-input space is partitioned into polyheral regions (See Fig. \ref{fig:PWADyna}).

Thus, by using the established tools for controlling PWA dynamics such as the parametric programming in \cite{mayne2003model,kvasnica2004multi}, one can compute the explicit solution of \eqref{eq:impMPC_MIP} (see e.g., Section 3 \cite{kvasnica2004multi}). In the subsequent parts, we will refer to this control implementation as explicit MPC (exMPC).


In the next part, the efficacy of the proposed settings will be analyzed via numerical simulations.
\section{Simulation study}
\label{sec:sim}

The following simulation tests demonstrate the validity of {the theoretical results} for various examples. Furthermore, the results yield insights into challenges and future research directions.

\subsection{Aircraft longitudinal dynamics}
\label{subsec:Aircraft}

\begin{figure}[htbp]
    \centering
\includegraphics[width=0.55\textwidth]{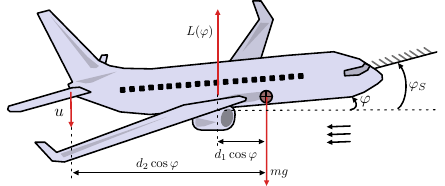}
    \caption{A longitudinal dynamic model of a civil aircraft (illustration based on \cite{nicotra2018explicit}).}
    \label{fig:aircraft}
\end{figure}

Consider the longitudinal model of a civil aircraft \cite{nicotra2018explicit}:
\begin{equation}
    \ddot \varphi =J^{-1}( -d_1L(\varphi) + u d_2 )\cos\varphi, \text{ s.t. } |u|\leq \overline u,\: \varphi \leq \varphi_S,
    \label{eq:aircraftModel}
\end{equation}
where $\varphi $ is the angle of attack. Changing this angle creates a lift force $L(\varphi) $ modeled as:
$L(\varphi) = l_0+l_1\varphi-l_3\varphi^3$ and the angle must be constrained below $\varphi_S =\sqrt{l_1/(3l_3)}$. The input $u$ is the elevator force with the maximum amplitude $\bar u$. $J$ is the longitudinal inertia, $d_1 $ and $d_2$ are the distances between the two airfoils (see Fig. \ref{fig:aircraft} for illustration).
The numerical values are given as $l_0=2.5\times 10^5,l_1 = 8.6\times 10^6, l_3 = 4.35\times 10^7$, $J=4.5\times10^6$Nm$^2$, $\bar u = 5\times 10^5$ N, $d_1 = 4 $m and $d_2  =42$m.



It is straightforward to linearize the system by setting the right-hand side as a new control variable $v$, namely, the equation \eqref{eq:aircraftModel} yields:
\be
\dfrac{\mathrm{d}}{\mathrm{d} t}\hspace{-0.35cm}\underbrace{\bbm \varphi \\ \dot \varphi \ebm }_{\bz = [z_1 \ z_2]^\top} = \underbrace{\bbm 0 & 1 \\ 0&  0\ebm}_{\bA} \bz +
\underbrace{\bbm 0  \\ 1\ebm}_{\bB} v,
\label{eq:linearizedAircraft}
\ee 
and the constraints become:
\be 
\big|\underbrace{d_2^{-1}(vJ\cos^{-1}z_1+d_1L(z_1)) }_{\Phi(\bz,v)}\big|\leq \bar u,\quad z_1\leq \varphi_S.
\label{eq:aircraftConstr}
\ee

Next, the mapping $\Phi(\bz,v)$ as in \eqref{eq:aircraftConstr} is approximated\footnote{For the training, we employ the \texttt{fitnet} method from MATLAB.} by a ReLU-ANN $\Phi_{nn}(\bz,v)$ structured as in \eqref{eq:ReLU_single} with three neurons. The numerical values are given as:
\be 
\begin{aligned}
    & \bW^1 = {\bbm 
\hphantom{-}6.9544  &  -0.7445\\
   -7.0939    & \hphantom{-}0.0354\\
    \hphantom{-}4.2001  & -0.0208\\
\ebm},\bW^2 =\bbm   -1.5659   &-1.2676&   \hphantom{-}2.1689
\ebm ,\\
& \bb^1 = [   14.9468 \   1.4271   \ 0.8521
]^\top, \bb^2 =    23.6044.
\end{aligned}
\label{eq:ANN_aircraft_parameters}
\ee

\begin{thinhnote}
For demonstration, we estimate the approximation error using Propositions \ref{prop:gridbasedError} and \ref{prop:error_esti} in this example.

\underline{$\bullet$ Applying Proposition \ref{prop:gridbasedError}:} 

The ingredients for the analysis therein include the Lipchitz constants of the linearizing mapping, the PWA map from the ANN, and the upper-bound of the granularity function over the grid of samples.

    First, the Lipschitz constant of the linearizing mapping $\Phi$ as in \eqref{eq:aircraftConstr} can be estimated as follows.
    Suppose that the examined workspace is limited as
\be
\mc Z = \{(z_1,v): |z_1| \leq{\bar{\varphi}}, |v|\leq \bar v
\}. \label{eq:examinedDomain}
\ee 
    Then, consider $(z_1,v) $, $(\tilde z_1, \tilde v)\in\mathcal Z$
    :
    \be 
\begin{aligned}
\left|\Phi\left(z_1, v\right)-\Phi\left(\tilde z_1, \tilde v\right)\right| & \leq\left|\Phi\left(z_1, v\right)-\Phi\left(z_1, \tilde v\right)\right|
+\left|\Phi\left(z_1, \tilde v\right)-\Phi(\tilde z_1, \tilde v)\right| \\
& \leq a\left|v-\tilde v\right|+b\left|z_1-\tilde z_1\right|  \leq\underbrace{\sqrt{a^2+b^2}}_{\gamma_\Phi} \left\|\bbm z_1\\v \ebm-\bbm \tilde z_1\\\tilde v \ebm\right\|_2, 
\end{aligned}
\label{eq:Lipschitz_Phi}
    \ee 
    where $a, b $ are the two constants bounding the deviation in each coordinate as:
\be 
\begin{cases}
    \left|\Phi\left(z_1, v\right)-\Phi\left(\tilde z_1, \tilde v\right)\right| & \leq \underbrace{\dfrac{J}{d_2\cos{\bar{\varphi}}}}_{a}\left|v-\tilde v\right|, \\
    \left|\Phi\left(z_1, \tilde v\right)-\Phi(\tilde z_1, \tilde v)\right|& \leq \underbrace{\dfrac{1}{d_2}\left(\dfrac{\bar v J}{\cos^2{\bar{\varphi}}}+d_1\left(l_1+3 l_3 {\bar{\varphi}}^2\right)\right)}_b\left|z_1-\tilde z_1\right|  .\rule{0pt}{20pt}
\end{cases} \label{eq:walkingwalking1}
\ee 
    Thus $\gamma_\Phi$ in \eqref{eq:Lipschitz_Phi} is the Lipschitz constant of the map.

    Second, with the notation $\bzeta = \bbm z_1 \\ v\ebm $ as in Proposition \ref{prop:gridbasedError}, the PWA map $\Phi_{nn}(\bzeta)$ given in \eqref{eq:ANN_aircraft_parameters} can be described as:
    \be 
\Phi_{nn}(\bzeta)= \begin{cases}
    \boldsymbol{F}_1\bzeta + f_1, &\text{ if } \bzeta\in \mc C_1,
    \\
    \boldsymbol{F}_2\bzeta + f_2, &\text{ if } \bzeta\in \mc C_2,
    \\
    \boldsymbol{F}_3\bzeta + f_3, &\text{ if } \bzeta\in \mc C_3,
\end{cases} \label{eq:threeCellsForLip}
    \ee 
with $F_j,f_j,\mc C_j$ enumerated via 
Algorithm \ref{algo:cellEnum}. The numerical values can be found in the file \texttt{u\_aircraft.mat} in our GitLab repository: \url{https://gitlab.com/huuthinh.do0421/constrained-control-of-flat-systems-with-relu-ann}. Then, the Lipschitz constant of the continuous PWA map can be computed as \cite{fabiani2024robust}:
\be 
\gamma_{nn} = \max_{i\in\{1,2,3\}} \|\boldsymbol{F}_i\|. \label{eq:LipPWA}
\ee 

Thus, via the triangle inequality, one can derive the Lipschitz constant of the approximation error $ \boldsymbol{\varepsilon}(\bzeta) = \Phi(\bzeta) - \Phi_{nn}(\bzeta)$ in \eqref{eq:lipschitz_approx_err} as:
\be 
|\boldsymbol{\varepsilon}(\bzeta_a) - \boldsymbol{\varepsilon}(\bzeta_b) |\leq 
(\gamma_\Phi + \gamma_{nn}) \|\boldsymbol{\varepsilon}(\bzeta_a) - \boldsymbol{\varepsilon}(\bzeta_b) \| \triangleq  \gamma_\varepsilon\|\boldsymbol{\varepsilon}(\bzeta_a) - \boldsymbol{\varepsilon}(\bzeta_b) \| .
\label{eq:WhyDidIputthis}
\ee 

Next, for this particular example, we choose a uniform grid $\mc Z_g$ over the two axes $z_1$ and $v$ as follows.

Let $\delta_z>0,\delta_v>0$ be the fixed step size in $z_1$ and $v$ directions, respectively. The uniform grid is defined as:
\be 
\mc Z_g = \{\bzeta = [i\delta_z,\ j\delta_v]^\top:\  |i\delta_z|\leq \bar\varphi, |j\delta_v|\leq \bar v,\  i,j\in\mathbb Z\}. \label{eq:gridDeltas}
\ee 
The elements of the set can be enumerated via the \texttt{colon} operator in MATLAB.

Finally, with the choice of such a uniform grid, the supremum of the granularity function yields:
\be 
\bar \rho = \sup \rho(\bzeta) =\sqrt{\left(\dfrac{\delta_z}{2}\right)^2 + \left(\dfrac{\delta_v}{2}\right)^2}. \label{eq:gridSize}
\ee 

Then, over the grid $\mc Z_g$, let $\tilde\epsilon = \displaystyle\max_{\bzeta_g\in \mc Z_g} |\Phi(\bzeta_g) - \Phi_{nn}(\bzeta_g)|$ as in \eqref{eq:gridBounds}, the upper bound of the approximation error can be given as:
\be 
|\Phi(\bzeta) - \Phi_{nn}(\bzeta)| \leq \tilde\epsilon + \gamma_\varepsilon\bar\rho \triangleq \bar\varepsilon.\label{eq:upperAirplaneBound}
\ee

With all those ingredients, the numerical values for the example are given as $\gamma_{\Phi}=29.42, \gamma_{nn} =7.29, \gamma_\varepsilon = 36.71$.
These values give us the estimation of how dense the grid $\mc Z_g$ should be. More specifically, suppose that we want $|\tilde\epsilon - \bar\epsilon|\leq\Delta_\epsilon$, then one needs
$$\bar\rho \leq \Delta_\epsilon/\gamma_\varepsilon.$$
Considering the maximum input value is $ \bar u =4$, we choose $\Delta_\epsilon = 0.025$, thus the requirement for the grid is $\bar\rho\leq 0.68\times 10^{-3}$, which can be satisfied by choosing $\delta_z=\delta_v = 0.9\times10^{-3}$.
The examined space $\mc Z$ are chosen with $\bar\varphi = 0.349$ rad (20 deg) and $\bar v = 5 $ (rad/s$^2$). 
This results in the grid $\mc Z_g$ with more than 8.6 million points. The verification of \eqref{eq:upperAirplaneBound} for all the points
is implemented with
Matlab 2021a, Intel Core, i5-10300H CPU @ 2.50GHz and 16GB
RAM. The computation time is around 51 seconds with the approximation error of $\epsilon=0.1897$.

\underline{$\bullet$ Applying Proposition \ref{prop:error_esti}:} 

The ingredients include the Lipschitz constants of the partial derivatives of $\Phi$:
\be
\begin{cases}
    \Phi_z(z_1,v) &= \dfrac{\partial \Phi}{\partial z_1 }(\bzeta) = \dfrac{1}{d_2}\left(d_1(l_1-3l_3 z_1^2) + Jv\dfrac{\sin z_1}{\cos^2 z_1} \right),\\
   \Phi_v(z_1,v) &=\dfrac{\partial \Phi}{\partial v}(\bzeta) = \dfrac{J\cos z_1}{d_2} \rule{0pt}{18pt}.
\end{cases} 
\label{eq:partialPhi}
\ee 

Similar to the previous computation, con, we can compute the Lipschitz bounds by showing:
\be 
\begin{cases}
    |\Phi_z(z_1,v) - \Phi_z(\tilde z_1, \tilde v)| & \leq C_z \left\|\bbm z_1\\v \ebm-\bbm \tilde z_1\\\tilde v \ebm\right\|_1, \\
      |\Phi_v(z_1,v) - \Phi_v(\tilde z_1, \tilde v)| & \leq C_v \left\|\bbm z_1\\v \ebm-\bbm \tilde z_1\\\tilde v \ebm\right\|_1 .\rule{0pt}{25pt}
\end{cases}
\label{eq:LipBounddPhi}
\ee 
Thus, $C_\zeta $ in \eqref{eq:SmoothAssumption} can be computed as $C_\zeta = \max\{C_z,C_v\}$ where

\be 
\begin{aligned}
C_z &= \max\left\{
\dfrac{J\sin \bar\varphi}{d_2\cos^2 \bar\varphi},    \frac{6 d_1l_3\bar\varphi}{d_2}+J \bar v
  \left(\dfrac{1}{\cos^2\bar\varphi}  + \dfrac{2\sin\bar\varphi}{\cos^4\bar\varphi}\right)\right\}, \\
    C_v &=  \dfrac{J}{d_2\cos^2\bar\varphi}.
\end{aligned} \label{eq:CzCv}
\ee 
 For brevity, the derivation of $C_z,C_v$ will be given in Appendix \ref{app:L_bound}.

The corresponding numerical values are $C_z = 538.9626$, $C_v = 1.2134$,  hence obtaining, $C_\zeta = C_v$. 

Next, to linearize the map, for each cell, we choose the linearization points at the cells' centers $\bzeta_{e^i}$ with the smallest radius $r_{e^i}$ such that the cell is contained inside the 1-norm ball centered at $\bzeta_{e^i}$, The centers and the radii are illustrated as the red points and the black arrows in Figure \ref{fig:Taylor_NN}, respectively. Their numerical values are given in Table \ref{tab:cellsTayor}. Consequently, the upper bound for the approximation error can be computed for each cell as $\bar \epsilon_T + \bar\epsilon_H$. 

Regarding the result, it is evident that the bound $\epsilon$ estimated from Proposition \ref{prop:error_esti}  exceeds the admissible control input and is therefore unusable.
We therefore rely on Proposition \ref{prop:gridbasedError} for the simulations. It can further be seen that Proposition \ref{prop:gridbasedError} is computationally heavy but conceptually straightforward: by refining the grid, the estimated bound can be made arbitrarily close to the true value. In contrast, Proposition \ref{prop:error_esti} offers an analytic, formula-based bound determined once the derivatives are known, but such bounds are typically local and may be overly loose.

\begin{table}[htbp]
    \centering
    \begin{tabular}{|c|c|c|c|}\hline
        Cell & $\mc C_1$ & $\mc C_2$ &$\mc C_3$\\  \hline
        $r_{e^i}$ &    4.9177&   3.6495&     3.6457\\ \hline
        $\bzeta_{e^i}$ & $[-0.0019   \ \  -0.2092]^\top$ & $[ -0.2732 \ \     0.9732]^\top$ & $[0.2713   \ \  -1.3966]^\top$\\ \hline
        $\bar\epsilon_T = C_\zeta r_{e^i}/2$ in \eqref{eq:ErrorTaylorBound} & 1325.2&  983.5&    982.5\\ \hline
       $\bar\epsilon_H$ in \eqref{eq:ErrorPWA_Taylor}& 0.2006&  0.1365&  0.1379\\ \hline
    \end{tabular}
    \caption{\thinh{Numerical results of the Taylor approximation of the linearizing mapping for the aircraft model.}}
    \label{tab:cellsTayor}
\end{table}

\begin{figure}[htbp]
    \centering
    \includegraphics[width=0.985\linewidth]{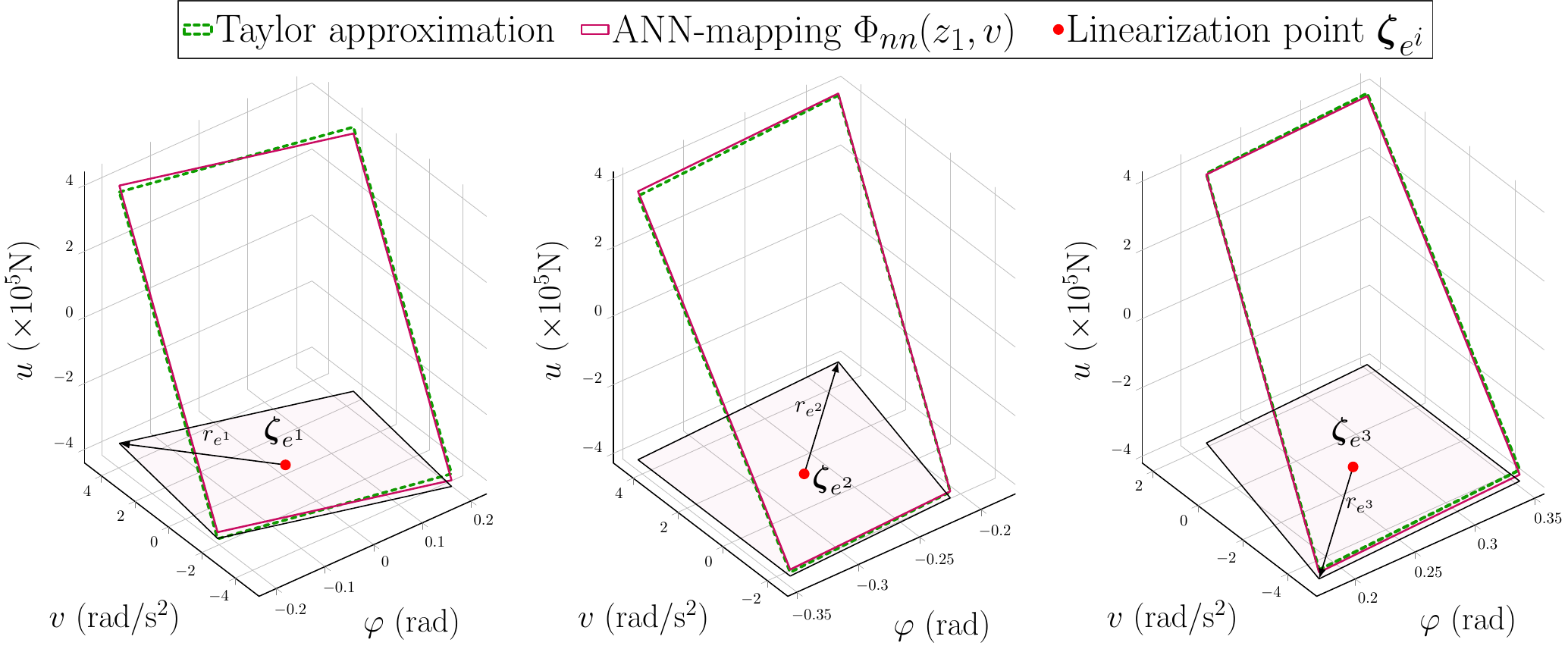}
    \caption{\thinh{Taylor approximation of the nonlinear mapping $\Phi(z_1,v) $ and the ANN approximation $\Phi_{nn}(z_1,v)$ for the aircraft model.}}
    \label{fig:Taylor_NN}
\end{figure}

\end{thinhnote}

\begin{figure}[htbp]
    \centering
    \includegraphics[width=0.725\linewidth]{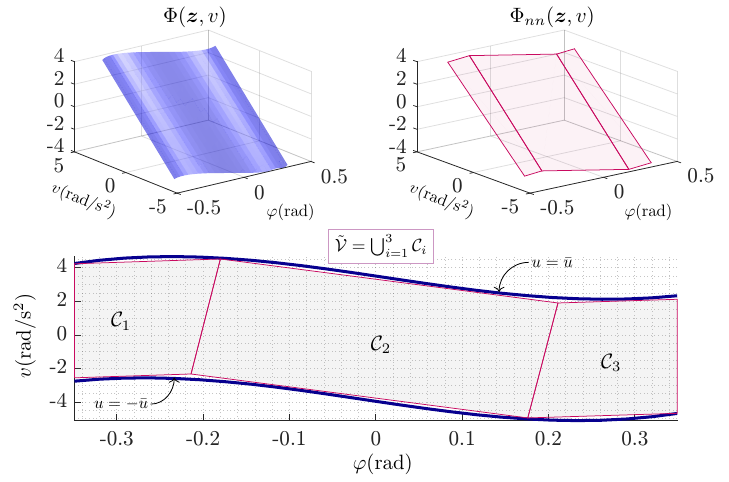}
    \caption{The linearizing mapping $\Phi(\bz,v)$ in \eqref{eq:aircraftConstr}, its approximation $\Phi_{nn}(\bz,v)$ and the corresponding polytopic cells.}
    \label{fig:cellsAircraft}
\end{figure}
\begin{figure}[htbp]
    \centering
    \includegraphics[width=0.8\linewidth]{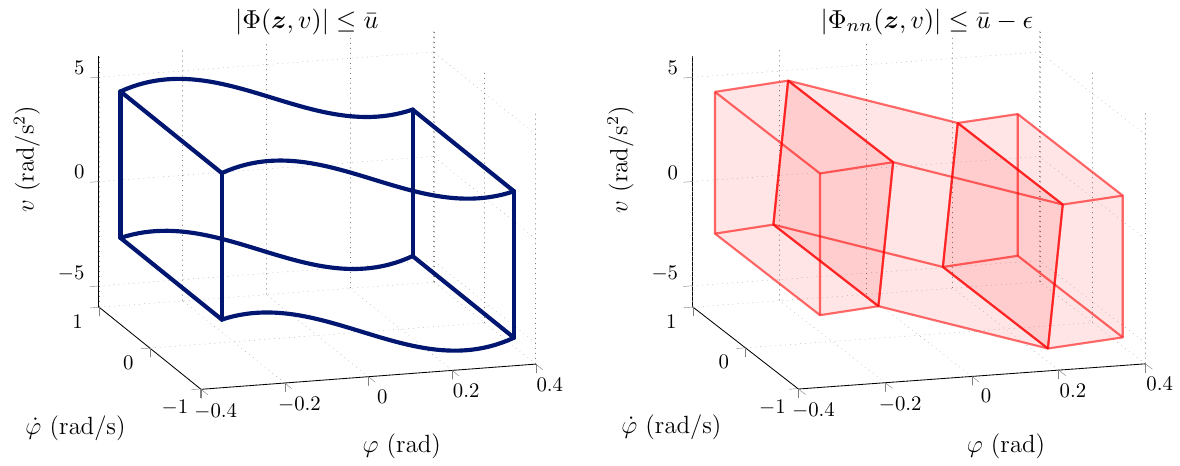}
    \caption{Convoluted constraint set (left) and its ANN-based approximation (right).}
    \label{fig:AirCr_true_approx_sets}
\end{figure}

With the estimated bound, the feasible domain can be described as a union of three polytopic cells: 
\be \tilde{\mc V} = \cup_{i=1}^3\mc C_i. \label{eq:unionAircraft}\ee
The mapping, its approximation and the set $\tilde{\mc V}$ in \eqref{eq:unionAircraft} are depicted in Fig. \ref{fig:cellsAircraft} and \ref{fig:AirCr_true_approx_sets}.

With the characterized constraint set $\tilde{\mc V}$
in the form of \eqref{eq:Vtilde_Hspace}, the previously presented CLF-based and MPC settings will be implemented for the linear system \eqref{eq:linearizedAircraft} subject to the constraints:
\be 
[\bz \ v]^\top\in \tilde{\mc V} \text{ as in \eqref{eq:unionAircraft} and } z_1 \leq \varphi_S.
\ee 

\begin{figure}[htbp]
    \centering
    \resizebox{0.95\linewidth}{!}{\definecolor{impc}{rgb}{ 0   , 0.4470   , 0.7410}
\definecolor{lcss}{rgb}{0.8500 ,   0.3250 ,   0.0980}
\definecolor{exMPC}{rgb}{0.9290  ,  0.6940  ,  0.1250}

\definecolor{CLF}{rgb}{0.4940   , 0.1840  ,  0.5560}
\def\ra{1.6}

\definecolor{blacktext}{HTML}{04002b}
\begin{tikzpicture}
    \draw (0,0) node {\includegraphics[scale=0.5]{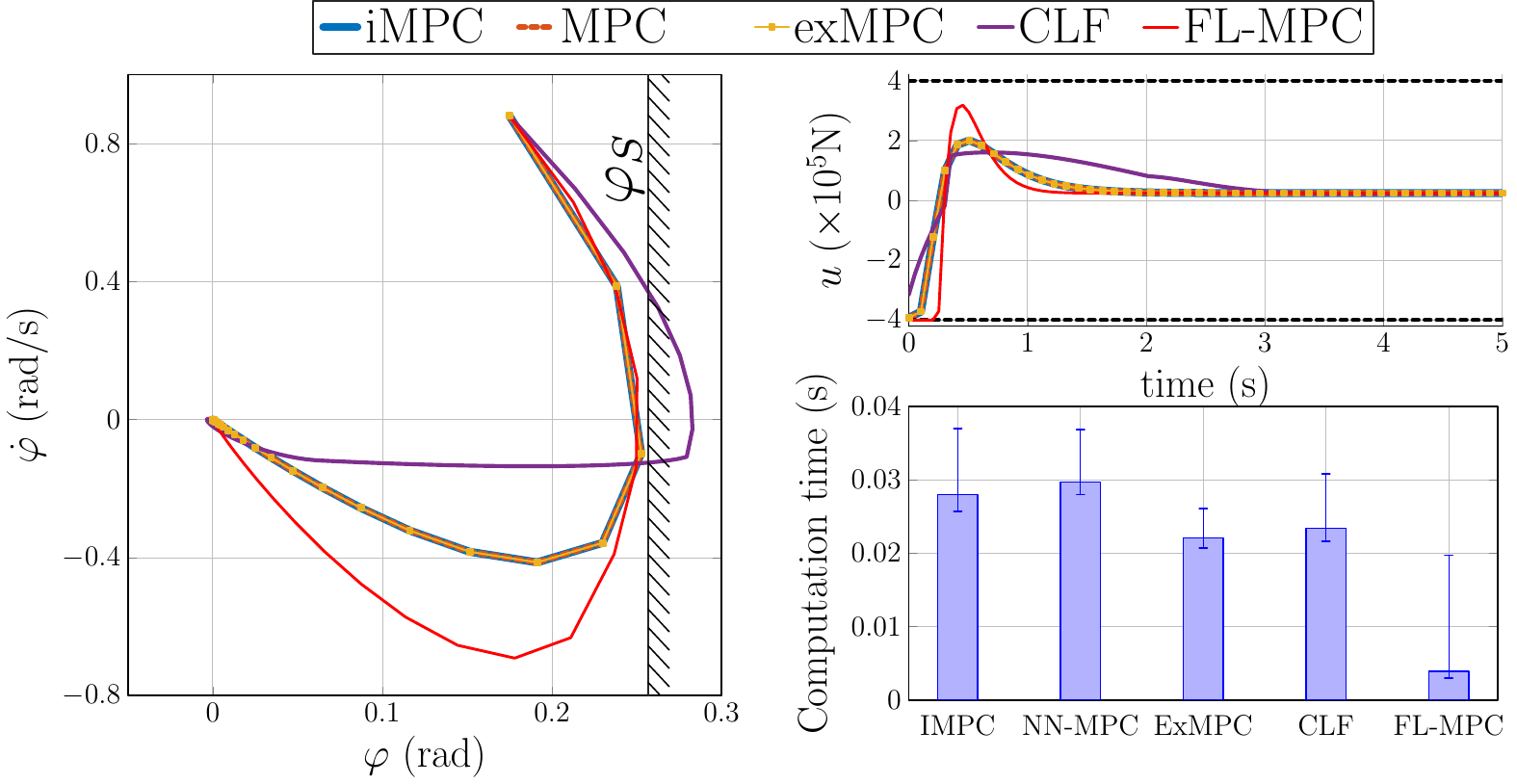}};
\draw (-0.35, 3.2) node {\cite{do2024_reluANN}};
    










    \draw (4.85,2.55) node [ scale= 0.85] {$u = \bar u $}; 
    \draw (4.85,0.8) node [ scale= 0.85] {$u = -\bar u $}; 
\end{tikzpicture}}
    \caption{Implementation of the presented settings.}
    \label{fig:compareMethods}
\end{figure}

For the CLF as in \eqref{eq:CLF-general-MIP}, with $\bP$ computed via \eqref{eq:LMI}, $\bP = \bbms0.1430   & 0.1932\\
    0.1932  &  0.6378\ebms $ with $\gamma = 0.05$. For the desired controller, we adopt the infinite horizon feedback gain $v_d(\bz) = K\bz = -[3.16   \, \ 2.55]\bz $.
With the MPC setting in \eqref{eq:impMPC_MIP}, the weights are set as $\bR = 0.005$ and $ \bQ = \bbms 20 & 1 \\ 1 &0.5\ebms$. All the ``big-M'' variables $M_j$ in \eqref{eq:CLF-general-MIP} and \eqref{eq:predictedConstraint} are set as $M_j=5000$. The sampling time for the MPC is chosen $T_s = 0.1$s and the prediction model is the forth-order Runge-Kutta discretization of \eqref{eq:linearizedAircraft}. The prediction horizon is $N_p = 5$.
The simulation with the implicit MPC and explicit MPC from Section \ref{subsec:MPCForm} are noted as iMPC and exMPC, respectively. 
For validation, we implement the ANN-based MPC setting from \cite{do2024_reluANN} where a binary variable is assigned to each neuron of the network. The tuning is chosen identical to the presented values. For implementation, the MIPs will be solved with the CPLEX solver. The simulation results of the four control strategies are given in Fig. \ref{fig:compareMethods}.

\begin{thinhnote}

For comparison, we implement a common FL-based MPC (FL-MPC) scheme for model \eqref{eq:linearizedAircraft}, which enforces the state constraint along the prediction horizon, while constraining only the first input to be applied. The online optimization problem for FL-MPC is given by:


\begin{subequations}
\label{eq:FLMPC}
\begin{align}
&\min_{\bv(k|\cdot)}\sum_{i=0}^{N_p-1}\|\bz(k|i) \|_{\bQ}^2 + \|\bv(k|i)\|_{\bR}^2  \label{eq:FLMPCa}\\
    &\bz(k|i+1) = \bA_d \bz(k|i) + \bB_d\bv(k|i), \ 
\label{eq:FLMPCb} \\
&\bz(k|0) = \bz_0, , \bz(k|i) \in \mc Z_s, i \in \{0,
..., N_p-1\}, \label{eq:stateConstrFLMPC}\\
&|\Phi(\bz(k|0),\bv(k|0))| \leq \bar u, \label{eq:firstInputConstr}
\end{align}
\end{subequations}
where the state constraint in this example is described as $  \mc Z_s = \{\bz: \bz_1 \leq \varphi_S\}$. 
By imposing \eqref{eq:stateConstrFLMPC}--\eqref{eq:firstInputConstr}, one can ensure that the state and input constraints will be enforced.
With the setting established, we proceed to the results.
\end{thinhnote}

First, it can be seen that both the exMPC and iMPC yield the same output as the MPC solution from \cite{do2024_reluANN} despite the differences in their formulations, reaffirming the correctness of the cell enumeration procedure. Additionally, possessing stabilizing properties, the CLF-based controller ensures adherence to input constraints but does not guarantee satisfaction of state constraints. These MPCs settings, again show effectiveness in handling both state and input constraints as shown in Fig.
\ref{fig:compareMethods}.

Second, regarding the online complexity for this particular example, we see a slight reduction in the computation time (CT) from the exMPC. This is because the solution is derived in the form of look-up tables. It is interesting to note that, although the CLF-based control employs the least binary variables, the computation time online showed little difference with respect to the MPCs.

\begin{figure}[htbp]
    \centering
    \includegraphics[width=0.95\linewidth]{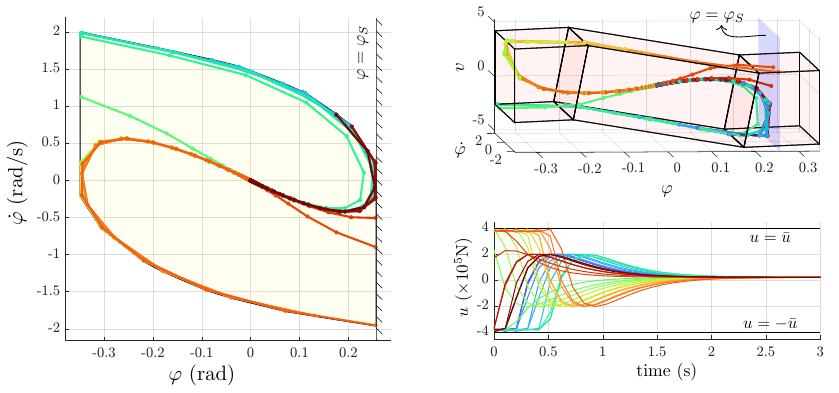}
    \caption{Simulation results with the explicit MPC solution from MPT3 \cite{kvasnica2004multi} {with different initial states}.}
    \label{fig:allvar_aircraft}
\end{figure}

With the explicit representation from the cell enumeration algorithm (Algorithm \ref{algo:cellEnum}), the explicit MPC solution further provides us an analysis on the feasibility. More specifically, as shown in Fig. \ref{fig:allvar_aircraft}, the set of initial points admitting a constrained stabilizing control under the MPC (the yellow set) has been identified {as the union of the critical regions.} In this manner, the domain of attraction of the nonlinear system \eqref{eq:aircraftModel} can be estimated for the stabilization problem. 

\begin{figure}[htbp]
    \centering
    \includegraphics[width=0.895\linewidth]{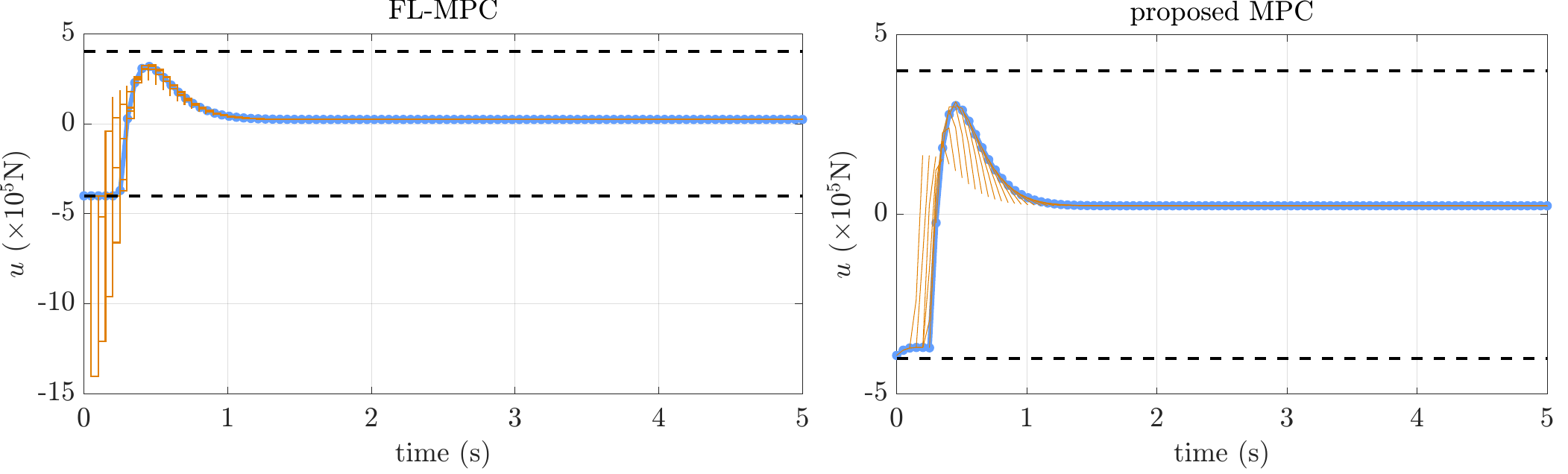}
    \caption{The applied input (blue) and the predicted input trajectory (yellow) of FL-MPC and the proposed MPC.}
    \label{fig:TwoFL_MPCs}
\end{figure}

\thinh{
Finally, compared with the baseline FL-MPC \eqref{eq:FLMPC}, which retains the simplicity of a quadratic program, the proposed setting is computationally more demanding. However, as illustrated in Fig.~\ref{fig:TwoFL_MPCs}, FL-MPC does not enforce the input constraint along the prediction horizon, whereas the proposed MPC guarantees that the forecasted trajectories (yellow) also respect this constraint. This condition is relevant when one needs to ensure recursive feasibility, i.e., the existence of a feasible control sequence not only for the current time instant but also in the future.
}

Next, we show that the MPC framework in Section \ref{subsec:MPCForm} can also be adapted for a trajectory tracking problem for an unmanned aerial vehicle (UAV) model.




\subsection{Planar UAV model}
\begin{figure}[htbp]
    \centering
    \includegraphics[width=0.35\linewidth]{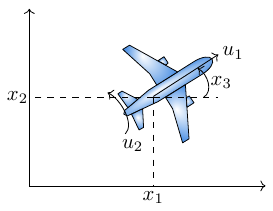}
    \caption{Three degree-of-freedom model of a fixed-wing UAV.}
    \label{fig:UAVsketch}
\end{figure}

In this part, let us consider the following model of a fixed-wing UAV \cite{stoican2017constrained}:
\begin{subequations}
\label{eq:UAVmodel}
\begin{align}
        &\dot x_1 = u_1 \cos x_3, 
        \dot x_2 = u_1 \sin x_3, 
        \dot x_3 = gu_2/u_1 \\
        &\text{s.t }0< \underline u_1 \leq|u_1 | \leq  \overline u_1, |u_2|\leq \overline u_2, \label{eq:UAVmodel_constraint}
    \end{align}
\end{subequations}
where $x_1, x_2$ denote the position of the plane and $x_3$ is the heading angle. The control input is the relative velocity and the tangent of the bank angle, denoted as $ u_1,u_2$ respectively (see Fig. \ref{fig:UAVsketch}). These inputs need to be constrained as in \eqref{eq:UAVmodel_constraint} with $\underline u_1 = 10$m/s, $\bar u_1 =26$m/s and $\bar u_2=    0.5774$. 

\begin{figure}[htbp]
    \centering
    \resizebox{0.7\linewidth}{!}{\definecolor{colorPWA}{rgb}{    0.2599,    0.2915  ,  0.7045
}
\begin{tikzpicture}
    \draw (0,0) node {\includegraphics[scale = 0.5]{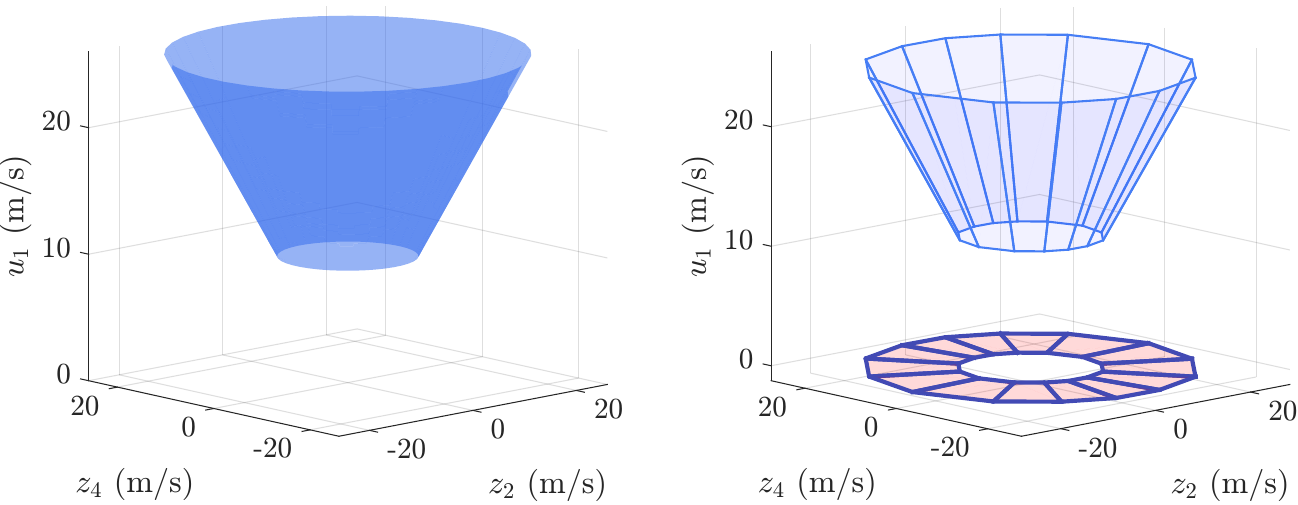}};
\draw (-2.5,2.35) node {$\Phi_1(\bz,\bv)$};
    \draw (3.25,2.35) node {$\Phi_{1,nn}(\bz,\bv)$};

    \draw (5.0,-0.25) node [scale=0.9, color = colorPWA] {$\mc V_{\Phi_1}$ in \eqref{eq:UAVVeloConstr}};

    \draw [->, color =colorPWA] (5,-0.425)  to [looseness = 1, out = -90, in =0,color = colorPWA] (5-0.7+0.37,-0.35-0.7);
\end{tikzpicture}}
    \caption{$\Phi_1(\bz,\bv)$ (left) and its approximation $\Phi_{1,nn}(\bz,\bv)$ (right).}
    \label{fig:UAVVeloConstr}
\end{figure}

It is known that the model \eqref{eq:UAVmodel} is differentially flat. Hence, by means of a variable change and input transformation, one can arrive to the linearized model:
\be
\label{eq:linearizedUAV_DI}
\dfrac{\mathrm{d}}{\mathrm{d} t}\underbrace{\bbm z_1 \\ z_2 \\ z_3 \\z_4\ebm }_{\bz } = \underbrace{\bbm 0 & 1 & 0 & 0 \\ 0&  0 & 0 &0 \\
 0 & 0 & 0 & 1 \\
 0&  0 & 0 &0 \ebm}_{\bA} \bz +
\underbrace{\bbm 0  & 0 \\
0& 1 \\
 0  & 0 \\
0& 1\ebm}_{\bB} \bv,
\ee 
with $z_1=x_2, z_2=x_2$. Correspondingly, the constraints become:
\begin{subnumcases}
{
\label{eq:UAV_constraint}
}
\underline u_1 \leq \Phi_1(\bz,\bv)= \sqrt{\dot z_1^2+\dot z_2^2} \leq \bar u_1 &\label{eq:VeloFlatConstr}\\
\Phi_2(\bz,\bv) = {\sqrt{ v_1^2+ v_2^2}} \leq \bar u_2{g}. \label{eq:bankangleFlatConstr}
\end{subnumcases}
For detailed derivation, see {Appendix \ref{ann:fixedWing}}. 
To characterize the constraints \eqref{eq:UAV_constraint}, we note that \eqref{eq:bankangleFlatConstr} describes $\bv$ being restricted in a two-dimensional ball. Hence, to deal with this constraint, we simply approximate the ball with a polytope by sampling its boundary:
\be
\bv \in \mc V_{\Phi_2} = \mathrm{co}\{\bv\in\R^2 : \bar u_2 g[\cos \theta, \sin \theta]^\top, \theta \in \mc N(0,2\pi,\ell)\}, \label{eq:TangentConstr}
\ee 
where $\mc N(0,2\pi,\ell)$ is the set of $\ell$ real number equally sampled from the interval $[0,2\pi]$ and $ \mathrm{co}\{\cdot\}$ denotes the convex hull operator.
Next, $\Phi_1(\bz,\bv)$ in \eqref{eq:VeloFlatConstr} will be approximated with a seven-node ReLU-ANN $\Phi_{1,nn}(\bz,\bv)$ of the form \eqref{eq:ReLU_single}. This network will be used to formulate the constraint:
\be 
\bbm \bz \\ \bv \ebm \in \mc V_{\Phi_1} = \{\bbm \bz & \bv \ebm^\top: \underline u_1 + \epsilon\leq \Phi_{1,nn}(\bz,\bv) \leq \bar u_1-\epsilon\}, \label{eq:UAVVeloConstr}
\ee
with {$\epsilon=0.981$} being the approximation error estimated. This set is then represented as in \eqref{eq:Vtilde_Hspace} via Algorithm \ref{algo:cellEnum}.
See Fig. \ref{fig:UAVVeloConstr} for the illustration of the approximated mapping and the corresponding union of polytopes $\mc V_{\Phi_1}$.

{Then, for trajectory tracking, we adapt the MPC in \eqref{eq:impMPC_MIP} with a new cost as:}
\begin{subequations}
\label{eq:impMPC_UAV}
\begin{align}
\min_{\bv(\cdot)}&\sum_{i=0}^{N_p-1}\|\bz(k|k+i) -\bz^{\mathrm{ref}}(k|k+i)\|_{\bQ}^2 + \|\bv(k|k+i)-\bv^{\mathrm{ref}}(k|k+i) \|_{\bR}^2 
\\
\text{s.t. }&
\bz(k|k) = \bz_0, \\
    &\bz(k|k+i+1) = \bA_d \bz(k|k+i) + \bB_d\bv(k|k+i), \ 
i \in \{0,
..., N_p-1\}\\
&\bv(k|k+i) \in \mc V_{\Phi_2} \text{ in \eqref{eq:TangentConstr}} , [\bz(k|k+i) ,\bv(k|k+i) ]^\top \in \tilde{\mc V}_{\Phi_1}\text{ in \eqref{eq:UAVVeloConstr}}, \label{eq:unionConditionMPC_UAV}
\end{align}
\end{subequations}
where the superscript ``$\mathrm{ref}$'' denotes the reference signal for the system to track.
\begin{figure}[htbp]
    \centering
    \includegraphics[width=0.92\textwidth]{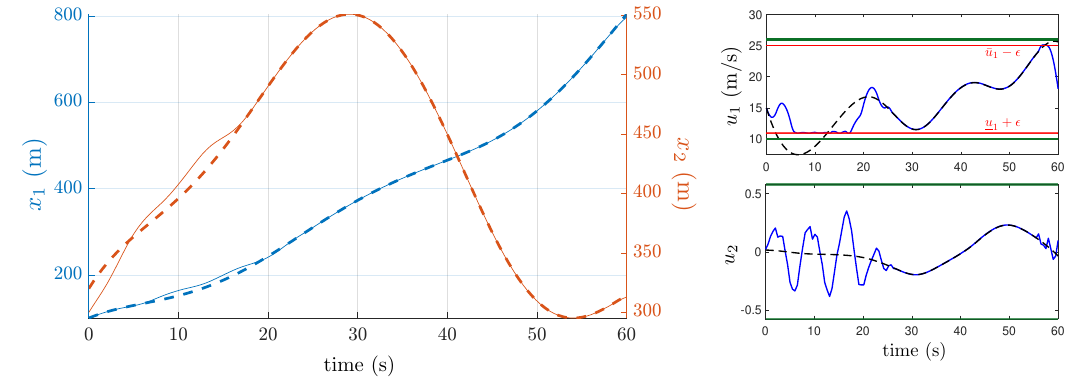}
    \caption{Trajectory tracking for the UAV model with constraint satisfaction (Dashed lines denote the reference signals while solid lines are the simulated trajectories with the controller; 
 the green lines mark the constraints for the inputs).}
    \label{fig:UAVSim}
\end{figure}

The simulated trajectory is given in Fig. \ref{fig:UAVSim}, confirming again the effectiveness of the ANN-based constraint characterization. In the final example, we report a particular case where the tracking is not satisfactory, although the constraints are successfully enforced. 



\subsection{PMSM model}

Let us consider the Permanent Magnet
Synchronous Motor (PMSM) model from \cite{vu2023port}:


\begin{subequations}
\label{eq:PMSMmodel}
\begin{align}
        \dot x_1 & = -\dfrac{R}{L} x_1 + \dfrac{1}{J_m}x_2x_3 + u_1 \\
     \dot x_2 & =   - \dfrac{1}{J_m}x_3(\Upsilon + x_1) - \dfrac{R}{L}x_2 + u_2 \\
     \dot x_3 & = \dfrac{\Upsilon}{L}x_2.
\end{align}
\end{subequations}
where $x_1, x_2$ are the stator magnetic fluxes and $x_3$ is the mechanical momentum. The input signals include the voltage $u_1,u_2$. {The mechanical inertia, the phase stator inductance, the phase resistance and the constant rotor magnetic flux are denoted as $J_m = 0.012$ kgm$^2$, $L_I=0.0038$mH, $R=0.225\Omega$ and $\Upsilon=0.17$Wb, respectively. }
With these values, the objective is to stabilize the system to the equilibrium point:
\be 
\bx_e=[0.0507,\ 0, \ 0.
1084]^\top \text{ and } \bu_e = [3, \ 1.9941]^\top .\label{eq:equiPointPMSM}
\ee 

The system \eqref{eq:PMSMmodel} is a differentially flat system. Indeed, with the flat output $z_1 = x_1, z_2=x_3$, one can linearize the model by changing the coordinate as:
\be
\label{eq:PMSM_coorchange}
\begin{cases}
    z_1 = x_1,&\\
    z_2 = x_3,&\\
    z_3 = \dot x_3 =\dfrac{\Upsilon}{L_I}x_2,
\end{cases}
\ee
with the input transformation:
\be
{\Phi(\bz,\bv)=\bbm v_1 + \dfrac{R}{L_I} z_1 - \dfrac{L_I}{J_m\Upsilon}z_2z_3, \\ 
 \dfrac{L_I}{\Upsilon}v_2 + \dfrac{1}{J_m}z_2(\Upsilon + z_1) + \dfrac{z_3}{\Upsilon}
\rule{0pt}{16 pt}\ebm} \label{eq:inputtransPMSM}
\ee


As a result, the dynamics in the linearizing space yields:
\be 
\dot \bz = \underbrace{\bbm 0 & 0 &0 \\
0 & 0 & 1 \\
0 &0 &0\ebm}_{\bA} \bz   + \underbrace{\bbm 
     1   &  0\\
     0   &  0 \\ 
     0   &  1 \\
\ebm}_{\bB} \bv.
\ee 

In accordance with the equilibrium point in \eqref{eq:equiPointPMSM}, we can compute the equilibrium point in the linearizing space from \eqref{eq:PMSM_coorchange}--\eqref{eq:inputtransPMSM}:
\be 
\bz_e = [0.0507\ 0.1084 \ 0]^\top, \bv_e = [0 \ 0 ]^\top. \label{eq:equiPointPMSM_flat}
\ee 

\begin{figure}[htbp]
    \centering
    \includegraphics[width=0.95\linewidth]{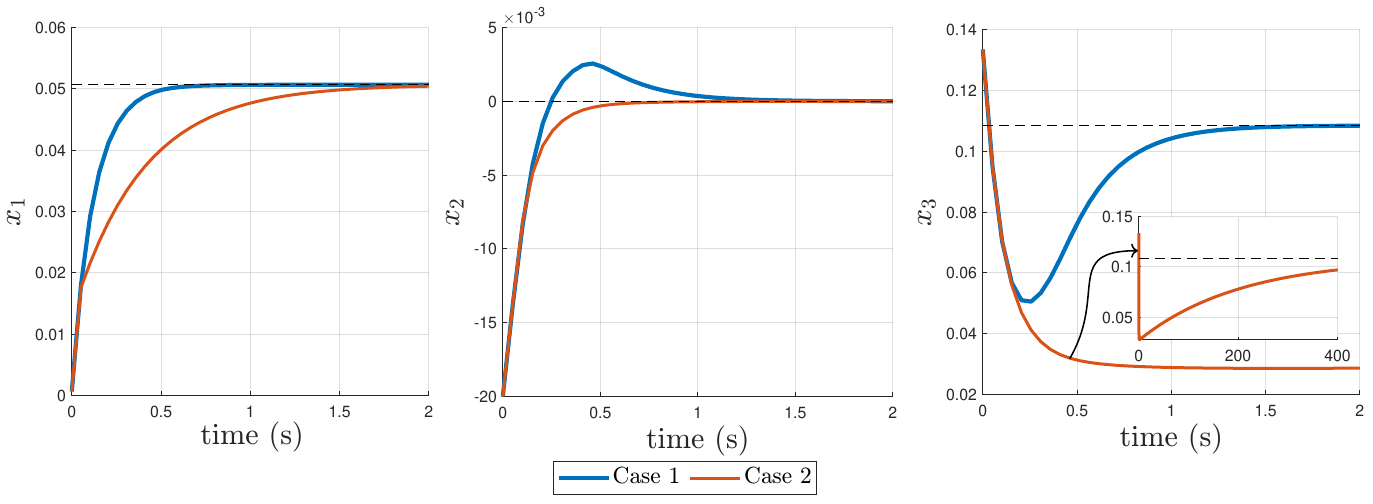}
    \caption{Undesired tracking with different tunings.}
    \label{fig:PMSMsim2cases}
\end{figure}

Similarly to the previous examples, we employ the MPC setting \eqref{eq:impMPC_MIP} to stabilize the system by shifting the origin to $\bz_e, \bv_e$ in  \eqref{eq:equiPointPMSM_flat}. The tuning choice is first given as $\bQ = \bbms 100 & 0 & 0 \\ 
0 &10 & 0 \\
0 & 0 &0.01\ebms$ and $N_p=5$. The sampling time is $T_s=0.05$s. To examine the stabilizing effect, we use two different values for $\bR$ in \eqref{eq:impMPC_MIP}: $\bR = 1\times 10^{-4}$ (Case 1) and $\bR = 1$ (Case 2), and the system's trajectory is given in Fig. \ref{fig:PMSMsim2cases}.

\thinh{The results reveal a significant weakness of the dynamics in \eqref{eq:impMPC_MIP}: the lack of a formal convergence guarantee. 
With Case 2, although achieved in a finite simulation time (around 400s for $x_3$), the convergence time is significantly larger with respect to Case 1, highlighting the importance of appropriate controller design and parameter selection.
}

\begin{table}[htbp]
    \centering
    \begin{tabular}{|l|c|c|c|}\hline
         & Aircraft & UAV & PMSM \\ \hline
       No. state \& input$(n,m)$ & (2,1) & (4,2) & (3,2)   \\ \hline
       No. cell as in \eqref{eq:Vtilde_Hspace}  & 3 & 14 & 10  \\\hline
       Prediction horizon $ N_p$ & 5 & 35  &   5\\\hline
        Avg. computation time (ms) & 28.04  &  186.29 &   34.71  \\\hline
    \end{tabular}
    \caption{Simulation specifications and results for the three examples.}
    \label{tab:Compttime}
\end{table}

\thinh{Furthermore, the computation times for the three examples are reported in Table \ref{tab:Compttime}, together with their number of states and inputs, the number of cells, and the prediction horizon. From these results, it can be seen that computational burden remains an open challenge. Even for the aircraft example with two states and one input, the average runtime is about 30 ms, which highlights the need for more efficient implementations, e.g., complexity reduction techniques for MIP \cite{prodan2015mixed}.}




\section{Conclusion}
\label{sec:concl}


This work {studies} the piecewise affine (PWA) representation of rectified linear unit (ReLU) neural networks to develop optimization-based control strategies for nonlinear differentially flat systems.  These strategies employ Control Lyapunov Functions (CLFs) and Model Predictive Control (MPC), integrated with flatness-induced feedback linearization.  Specifically, flatness provides a coordinate system in which the dynamics become linear yet the constraints are distorted. To handle this problem, by approximating the nonlinear constraints with ReLU 
neural networks and enumerating the polyhedral cells induced by the network's partitioning, a mixed-integer linear representation of the distorted constraints is obtained. This representation was followed by the integration to the CLF-based and MPC controllers. The simulation tests confirm the validity of the proposed setting, including the cell enumeration algorithm and the constraint satisfaction guarantee.
Briefly, with this representation of the approximated constraint and the linearized dynamics, it is hoped that the control design of nonlinear flat systems can be facilitated, from the offline design phase to the online implementation.
Future research will aim to reduce the computational overhead of the online controller and establish formal stability guarantees.

\backmatter



\bmhead{Acknowledgements}
This work received funding from the French government under the management of the National Research Agency (ANR) as part of the France 2030 program, reference ANR-23-IACL-0006.










\begin{appendices}

\section{Feedback linearization and constraint characterization for the fixed-wing UAV}\label{ann:fixedWing}

With the choice of flat output $z_1=x_1,z_2 = x_2$, one can linearize the model
\eqref{eq:UAVmodel_constraint} by setting the new variable system as:
\begin{subequations}
    \begin{align}
        x_1 &= z_1, \\
        x_2 &= z_2, \\
        x_3 &= \arctan \dfrac{\dot z_2}{\dot z_1} ,\\
        u_1 & = \sqrt{\dot z_1^2 + \dot z_2^2}  , \label{eq:VeloFlatrep}\\ 
          u_2 & = \dfrac{v_2\dot z_1 - v_2\dot z_2}{g\sqrt{\dot z_1^2 + \dot z_2^2}}.
    \end{align}
\end{subequations}
Consequently, one arrives to \eqref{eq:linearizedUAV_DI}. 

Regarding the constraint \eqref{eq:UAV_constraint}, \eqref{eq:VeloFlatConstr} can be found by simply replacing \eqref{eq:VeloFlatrep} into 
\eqref{eq:UAVmodel_constraint} while \eqref{eq:bankangleFlatConstr} is formulated by considering:
\be 
|u_2| = \left|\dfrac{v_2\dot z_1 - v_2\dot z_2}{g\sqrt{\dot z_1^2 + \dot z_2^2}}\right|\leq \dfrac{1}{g}\sqrt{(v_1^2+v_2^2)\left(\dfrac{\dot z_1^2 + \dot z_2^2}{\dot z_1^2 + \dot z_2^2}\right)} =\dfrac{\sqrt{v_1^2+v_2^2}}{g},
\ee 
where the first inequality follows from the Cauchy Schwarz inequality.
Hence $|u_2|\leq \bar u_2$ can be guaranteed by imposing:
\be
\dfrac{\sqrt{v_1^2+v_2^2}}{g} \leq \bar u_2,
\ee 
or \eqref{eq:bankangleFlatConstr}, completing the proof.

\begin{thinhnote}
\section{Derivation of the Lipschitz bounds for the aircraft model.}
\label{app:L_bound}

Consider $\Phi_z$ in \eqref{eq:partialPhi} with the domain of interest $\mc Z$ as in \eqref{eq:examinedDomain}. One has:
\be 
\begin{aligned}
    |\Phi_z(z_1,v) - \Phi_z(\tilde z_1,\tilde v)| \leq |\Phi_z(z_1,v) - \Phi_z( z_1,\tilde v)| 
    + |\Phi_z(z_1,\tilde v) - \Phi_z(\tilde z_1, \tilde v)|. 
\end{aligned}
\label{eq:Phi_z}
\ee 
The first term can be bounded as:
\be 
\begin{aligned}
    |\Phi_z(z_1,v) - \Phi_z( z_1,\tilde v)| &\leq \dfrac{J}{d_2} \dfrac{\sin z_1}{\cos^2 z_1} |v-\tilde v| \leq 
    \dfrac{J}{d_2} \dfrac{\sin \bar\varphi}{\cos^2 \bar\varphi} |v-\tilde v|.
\end{aligned}
\ee 
The second term yields:
\be 
\begin{aligned}
&|\Phi_z(z_1,\tilde v) - \Phi_z(\tilde z_1, \tilde v)| \leq \frac{1}{d_2}\left|-3 d_1l_3\left(z_1^2-\tilde z_1^2\right)+J \tilde v\left(
\dfrac{\sin z_1}{\cos^2 z_1}-
\dfrac{\sin \tilde z_1}{\cos^2  \tilde z_1}\right)\right| \\
& \leq 
 \frac{6 d_1l_3\bar\varphi}{d_2}|z_1-\tilde z_1|+\left|J \tilde v\left(
\dfrac{\sin z_1}{\cos^2 z_1}-
\dfrac{\sin \tilde z_1}{\cos^2  \tilde z_1}\right)\right| 
\\
&\leq 
 \frac{6 d_1l_3\bar\varphi}{d_2}|z_1-\tilde z_1|+J \bar v\left|
\dfrac{\sin z_1}{\cos^2 z_1}-
\dfrac{\sin \tilde z_1}{\cos^2  \tilde z_1}\right| 
\\
&\leq 
 \frac{6 d_1l_3\bar\varphi}{d_2}|z_1-\tilde z_1|+J \bar v\left|
\dfrac{\sin z_1 - \sin \tilde z_1}{\cos^2 z_1}+
\sin \tilde z_1\left(\dfrac{1}{\cos^2   z_1} - \dfrac{1}{\cos^2  \tilde z_1}\right)\right| 
\\
&\leq 
\left( \frac{6 d_1l_3\bar\varphi}{d_2}+J \bar v
  \left(\dfrac{1}{\cos^2\bar\varphi}  + \dfrac{2\sin\bar\varphi}{\cos^4\bar\varphi}\right)\right)
  |z_1-\tilde z_1|.
\end{aligned}
\label{eq:dPhi_z2}
\ee 

Thus, from \eqref{eq:Phi_z}--\eqref{eq:dPhi_z2}, one can derive $C_z$ in \eqref{eq:LipBounddPhi} as:
\be 
C_z = \max\left\{  \dfrac{J}{d_2} \dfrac{\sin \bar\varphi}{\cos^2 \bar\varphi},   \left( \frac{6 d_1l_3\bar\varphi}{d_2}+J \bar v
  \left(\dfrac{1}{\cos^2\bar\varphi}  + \dfrac{2\sin\bar\varphi}{\cos^4\bar\varphi}\right)\right)\right\}.
\ee

Next, consider $\Phi_v$ in \eqref{eq:partialPhi} one simply gets:
\be 
|\Phi_z(z_1,v) - \Phi_z(\tilde z_1,\tilde v) | \leq \dfrac{J}{d_2\cos^2\bar\varphi} |z_1 -\tilde z_1| \triangleq C_v |z_1 -\tilde z_1|.
\ee 

\section{Verification of the big-M constants}
\label{app:BigM}

\thinh{The verification for the big-$M$ number can be carried out as.}

\begin{thinhnote}
    For some polytope $\mc C=\{\bzeta: \Theta\bzeta\leq \btt\}$ consider the inclusion constraint in the big-M technique:
    \be 
    \label{eq:beta}
\Theta\bzeta\leq \btt + \beta M,
    \ee 
    with $\beta\in\{0;1\}.$
    The role of $M$ is to toggle the constraint between ``active'' and ``ignored'':

    \begin{equation*}
   \Theta\bzeta\leq \btt + \beta M
    \;\;\Rightarrow\;\;
    \begin{cases}
        \Theta\bzeta\leq \btt, & \beta=0,\\
       \Theta\bzeta\leq  \infty, & \beta=1.
    \end{cases}
\end{equation*}

In practice, $M$ cannot be set to $\infty$ or taken arbitrarily large. Instead, given a polytopic domain of interest $\mathcal Z$ in which the state evolves, $M$ is chosen so that $ \Theta\bzeta\leq \theta + \beta M$ is redundant for all $\bzeta\in\mathcal Z$, i.e.,
 \be 
\Theta_{[j,:]}\bzeta -  \btt_j\leq M,\ \forall \bzeta\in\mc Z, \ \forall j\in\{1,...,\mathrm{ len}(\btt)\}, \label{eq:redundant_1poly}
 \ee 
with $\mathrm{ len}(\btt)$ being the number of entries of $\btt$.
Then, as suggested in ~\cite{ioan2021mixed}, to satisfy \eqref{eq:redundant_1poly}, one can choose $M \geq M^\star$ by solving:
\be 
M^\star = \max_{j\in\{1,...,\mathrm{ len}(\btt)\}} \ \max_{\bzeta\in\mc Z}(\Theta_{[j,:]}\bzeta -  \btt_j), \label{eq:MinM}
\ee 
which exhibits the complexity of $\mathrm{ len}(\btt)$ linear programs. 
Geometrically, this corresponds to translating the edges of the polytope $\mc C$ just far enough so that they stay outside the region of interest $\mathcal Z$. Figure \ref{fig:ChooseM} illustrates the effect of $M^\star$ for the polytope $\mc C_2$ as in \eqref{eq:unionAircraft} with $\Theta_2=\begin{bsmallmatrix}
   7.212 &  -7.212   & 7.094   &-4.200 \\
    1.076  & -1.076  & -0.035  &  0.021
\end{bsmallmatrix}^\top$ $\btt_2 = [ 3.571\    4.049\    1.427\    0.852]^\top$ and $M^\star = 4.3247$.
\begin{figure}[htbp]
    \centering
    \includegraphics[width=0.75\linewidth]{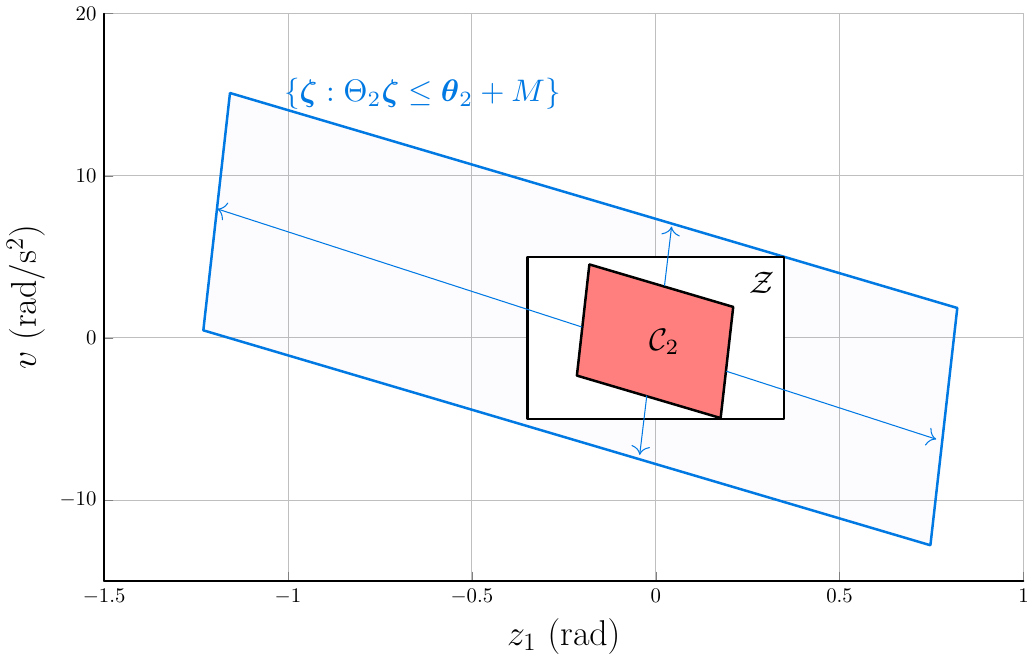}
    \caption{\thinh{While the constraint is active (i.e., $\beta=0$ in \eqref{eq:beta}), the resulting inequality:  $\Theta\bzeta\leq \theta$  describes the red region. Once $\beta=1$ (inactive/relaxed constraint), the selection of $M^\star = 4.3247$ inflates the set sufficiently large such that the remainder of the constraints is not affected.}}
    \label{fig:ChooseM}
\end{figure}

\end{thinhnote}

\end{thinhnote}




\end{appendices}


\bibliography{main}

\end{document}